\documentclass[iop,usenatbib,usegraphicx,border=0pt]{emulateapj}

\usepackage{natbib,graphicx}
\usepackage{epsfig}
\usepackage{ulem}
\usepackage{color}
\usepackage{graphics,graphicx}
\usepackage{times} 
\usepackage{amssymb}
\usepackage{amsmath}
\usepackage{url}
\usepackage{multirow}
\newif\ifAMStwofonts
\AMStwofontstrue

\def\aap{A\&A}                   
\def\apjs{ApJS}                  
\def\apjl{ApJ}                   

\def\xmm{{\it XMM-Newton~\/}}

\def\rxte{{\it RXTE~\/}}

\def\suzaku{{\it Suzaku}}

\def\bepposax{{\it BeppoSAX~\/}}

\def\epicmos1{{\it EPIC}{\rm-MOS1~\/}}
\def\epicmos2{{\it EPIC}{\rm-MOS2 ~\/}}
\def\epicmos{{\it EPIC}{\rm-MOS}}

\def\bepposax{{\it BeppoSAX}}

\def\xmm{{\it XMM-Newton}}

\def\rxte{{\it RXTE}}

\def\xspec{\hbox{\sc XSPEC}}
\def\xspecv{{\sc XSPEC}{\rm\thinspace v\thinspace 12.5.0}}

\def\heasoftv{\hbox{\rm HEASOFT\thinspace v6.12}}

\def\ks{\hbox{$\rm\thinspace ks$}}

\def\deg{$^{\circ}$}  

\def\pcmsq{\hbox{$\rm\thinspace cm^{-2}$}}

\def\ev{\hbox{$\rm\thinspace eV$}}
\def\kev{\hbox{$\rm\thinspace keV$}}

\def\ergcmps{\hbox{$\rm\thinspace erg~cm~s^{-1}$}}

\def\msun{\hbox{$\rm\thinspace M_{\odot}$}}

\def\rg{${\it r}_{\rm g}$}
\def\rin{${\it r}_{\rm in}$}

\def\ka{$K\alpha$}

\def\k{\hbox{\rm\thinspace K}}

\def\chisq{{\chi^{2}}}

\def\po{\rm{\small POWERLAW}}

\def\laor{\rm{\small LAOR}}
\def\gaussian{``Gaussian"}

\def\phabs{\rm{\small PHABS}}

\def\diskbb{\rm{\small DISKBB}}

\def\cflux{\rm{\small CFLUX\/}}

\def\reflionx{\rm{\small REFLIONX}}
\def\refbhb{\rm{\small REFBHB}}

\def\smedge{\rm{\small SMEDGE\/}}
\def\kdblur{\rm{\small KDBLUR}}
\def\kdblurf{\rm{\small KDBLURf}}

\def\pexriv{\rm{\small PEXRIV}}

\def\xspec{\hbox{\small XSPEC~\/}}
\def\xspecv{\hbox{\small XSPEC}\thinspace v12.7.0\thinspace}

\def\heasoftv{\hbox{\rm{\small HEASOFT}~v6.12\/}}

\def\grid25{\hbox{\rm{\small GRID25}}}

\def\pile_est{\hbox{\rm{\small PILE_EST}}}

\def\hxtback{\hbox{\rm{\small HXTBACK}}}

\def\hxtdead{\hbox{\rm{\small HXTDEAD}}}
\def\pcabackest{\hbox{\rm{\small PCABACKEST}}}
\def\xtefilt{\hbox{\rm{\small XTEFILT}}}

\def\j1118{\hbox{\rm XTE J1118+480}}
\def\j1749{\hbox{\rm J17497-2821}}

\def\j1752{\hbox{\rm XTE~J1752--223}}

\def\j{\hbox{\rm XTE~J1650--500}}

\shorttitle{Light-Bending and Spectral States}
\shortauthors{Reis et al.}

\begin{document}

\title{Evidence of Light-Bending Effects and its implication for spectral state transitions } 
\author{
R.~C.~Reis\altaffilmark{1},
J.~M.~Miller\altaffilmark{1},
M.~T.~Reynolds\altaffilmark{1},
A.~C.~Fabian\altaffilmark{2},
D.~J.~Walton\altaffilmark{2,3},
E.~Cackett\altaffilmark{4}, \&
J.~F.~Steiner\altaffilmark{2} 
}
\altaffiltext{1}{Dept. of Astronomy, University of Michigan, Ann Arbor, Michigan~48109, USA}
\altaffiltext{2}{Institute of Astronomy, University of Cambridge, Madingley Rd., Cambridge CB3 0HA, UK}
\altaffiltext{3}{Space Radiation Laboratory, California Institute of Technology, Pasadena, CA 91125, USA   }
\altaffiltext{4}{Department of Physics \& Astronomy, Wayne State University, 666 W. Hancock St, Detroit, MI 48201}

\begin{abstract}  

It has long been speculated that the nature of the hard X-ray corona may be an important second driver of black hole state transitions, in addition to the mass accretion rate through the disk.  However, a clear physical picture of coronal changes has not yet emerged.  We present results from a systematic analysis of Rossi X-ray Timing Explorer observations of the stellar mass black hole binary \j.  All spectra with significant hard X-ray detections were fit using a self-consistent, relativistically-blurred disk reflection model suited to high ionization regimes.  Importantly, we find evidence that both
the spectral and timing properties of black hole states may be partially driven by the height of the X-ray corona above the disk, and related changes in how gravitational light bending affects the corona--disk interaction.  Specifically, the evolution of the power-law, thermal disk, and relativistically--convolved reflection components in our spectral analysis indicate that: (1) the disk inner radius remains constant at \rin$=1.65\pm0.08~GM/c^2$ (consistent with values found for the ISCO of \j\ in other works) throughout the transition from the \textit{brighter} phases of the low-hard state to
the intermediate states (both the hard-intermediate and soft-intermediate), through to the soft state and back; (2) the ratio
between the observed reflected X-ray flux and power-law continuum (the ``reflection fraction'', $R$) increases sharply at the transition between the hard-intermediate and soft-intermediate states (``ballistic'' jets are sometimes launched at this transition); (3) both the frequency and coherence of the high-frequency quasi-periodic oscillations (QPOs) observed in \j\ increase with $R$.  We discuss our results in terms of black hole states and the nature of black hole
accretion flows across the mass scale.

\end{abstract}

\begin{keywords} { X-rays:  accretion -- accretion disks -- black hole physics -- line: profiles -- relativistic processes -- X-ray binaries -- X-ray Individuals: \j }\end{keywords}

\section{Introduction}
\label{introduction}

The majority of stellar-mass black holes residing in low mass binary systems spend most of their lives in quiescence. This explains the empirical fact that out of tens of thousands of such systems predicted to exist throughout our Galaxy \citep[e.g.][]{Yungelson2006}, only about 50 have been discovered. For a number of such systems, it is well established that the outburst -- which resulted in their discovery -- evolves through a number of spectral states characterised by the relative strength of their thermal and non-thermal X-ray emission and with possible differences in the accretion  geometry and reflection
 attributes. These active states can be roughly separated into four semi-distinct states which are phenomenologically described below and extensively discussed in \citet{Remillard06} and \citet{Bellonibook2010}. At the outset of the outburst, the system goes through what has been traditionally dubbed the low-hard state (LHS) where the X-ray spectrum is dominated by a non--thermal component often simply described by a power-law (photon index $\Gamma$ between $\sim1.4-2$)  spectrum with relatively low luminosity ($\sim 0.05 L_{\rm Edd} $ and and exponential cut-off at $\sim100$\kev).   The  energy spectrum in the LHS peaks near $\sim 100$\kev\ and often we also see the  weak presence of a thermal component (contributing  $<20\%$ of the total 2-20\kev flux)  with a temperature below $\sim0.5$\kev\  produced by the accretion disk \citep[see e.g.][and references therein]{Reynold2011swift, reislhs}.

As the luminosity increases, the spectrum moves through the intermediate state (IS)  where the 2-10\kev\ flux    is  typically a factor of $\sim4$ times higher than that of the LHS. Here, the soft ($\Gamma = 2-3$) power-law tail coexists with a strong thermal component. Recently, the intermediate state has begun to be subdivided into an early Hard-Intermediate State (HIS) and a later Soft-Intermediate state (SIS)  just prior to a transition into the canonical High-Soft or thermal state, where  the X-ray flux is dominated ($>75\% $ of the total 2-20\kev flux) by the thermal radiation from the inner accretion disk having an effective temperature of  $\sim1$\kev. In this final state of the outburst, the system usually emits with luminosities $>0.1 L_{\rm Edd}$ and the  power-law component is
both weak (less than 25\% of the total 2-20\kev\ flux) and steep ($\Gamma = 2-3$).  Following the HSS, the system often returns to the LHS and subsequently goes back to quiescence were it can remain indefinitely or in some cases for a few years before this cycle restarts.

The hard X-ray emission predominant in the LHS has long been linked to inverse Compton scattering of the soft thermal disk photons by a population of hot  ($\sim10^9\k$) electrons in a cloud of optically thin, ionised gas or ``corona" surrounding the inner parts of the accretion disk \citep{Shapiro1976,SunyaevTitarchuk1980}.  Under the common assumption that the radio emission observed to originate from stellar mass black holes, is directly related to the presence of a jet, it is believed that all such systems, either in the LHS or in transition, launch a collimated  outflow \citep[e.g.][]{Fender2001jets, fenderetal04, Fender091}. The fact that these persistent jets are observed only in the LHS suggests that the jet is linked to the corona, with claims that the corona in the LHS is indeed the launching point of persistent  jets \citep[see e.g.][]{Markoff05}. The connection between the radio (jet) and X-ray flux for both stellar-mass and supermassive black holes \citep{fundamentalplane,GalloFenderPooley2003,fundamentalplane2}, often referred to as ``the fundamental plane" of black hole accretion, suggests an intimate connection between the corona and radio-jets \citep[see e.g.][]{miller2012cyg}.  Whether state transitions are driven by intrinsic changes in $\dot{m}$, physical changes in the disk,  disk-corona, radio jet or a combination of all these factors is a matter of much debate.

\subsection{Reprocessed X-rays: Reflection }
The existence of a hard X-ray source -- the corona -- also adds further complexities to the various spectral states. The reprocessing of these hard X-rays by the relatively cold accretion disk in all active states results in a number of ``reflection features" consisting of discrete atomic features together with a ``Compton-hump" peaking at approximately 30\kev. The high fluorescent yield -- and relatively high cosmic abundance -- of iron often results  in a particularly strong feature at $\sim 6-7$\kev\ \citep[see e.g.][for a recent review of  ``reflection" in black holes]{Fabianross2010}.

The strong irradiation of the black hole accretion disk by the coronal photons likely causes the surface layers to be photoionised. \citet{rossfabian1993} investigated the effect of allowing the gas constituting the top layers of the accretion disk to ionise, and the authors went on to compute reflection spectra for different ionization levels. A number of similar studies of reflection from ionised matter have been conducted since \citep{MattFabianRoss1993, MattFabianRoss1996, rossfabianyoung99,NayakshinKazanasKallman2000, NayakshinKallman2001, cdid2001, GarciaKallman2010, GarciaKallman2011}. These studies demonstrate that  the reflection spectrum expected from a black hole depends strongly on the level of ionization of the surface layers of the disk. This can be quantified for a constant density gas by the ionization parameter \begin{equation}\xi  = \frac{L_{\rm x}}{nd^2},
                                                                                                                                                                                                                                                                                                                                                                                                                                                                                                                                                                                                                                                                                                                                                                                                                                     \end{equation}
where $L_{\rm x}$ is the ionising luminosity of the source, $d$ is the distance between the disk and the source, and $n$ is the density of the disk. Thus an increase in $\xi$, either by increasing the illuminating flux, decreasing the density or distance between the X-ray source and the disk, will cause the gas in the disk to become more ionised. \citet{MattFabianRoss1993, MattFabianRoss1996} split the behaviour of the reflection spectrum into four main regimes depending on the value of $\xi$.

For low ionization parameter ($\xi < 100~\ergcmps$),   the material is only weakly ionised and the reflection spectrum resembles that arising from  ``cold'' matter, with a prominent iron line at 6.4\kev, and strong absorption below $\approx 10$\kev. There is only a weak contribution from the backscattered continuum at $\approx 6$\kev\ and a weak iron K absorption edge at 7.1\kev. As the disk becomes more ionised ($100 < \xi < 500~\ergcmps$) the system reaches  intermediate ionization range where Fe has lost all of its M-shell ($n=3$) electrons and thus exists in the form of FeXVII--FeXXIII with a vacancy in the L-shell of the ion. Due to this vacancy, the L-shell can absorb the \ka\ line photons and thus effectively trap the escaping photon. This resonance trapping is only terminated when an Auger electron is emitted. This second  ionization regime is therefore characterised by a very weak iron line and a moderate iron absorption edge.
 
 As the gas becomes more ionised ($500 < \xi < 5000~\ergcmps$) all low-$Z$ metals are found in their hydrogenic form and  the soft reflection spectrum has only weak spectral features. Iron is found mostly in its hydrogen or helium like forms (FeXXVI or FeXXV respectively) and due to the lack of at least 2 electrons in the L-shell (i.e. not having a full 2s sub-shell), Auger de-excitation cannot occur. The result is strong emission from ``hot'' \ka\  FeXXV and FeXXVI at 6.67 and 6.97\kev\ respectively and the corresponding absorption edges at approximately 8.85 and 9.28\kev\ respectively. Finally, when $\xi \gg 5000~\ergcmps$, the disk is  highly ionised and there is a distinct absence of any atomic features.

A further complication arises in the reflection spectra of stellar mass black holes due to the fact that in these systems the gas in the accretion disk is inherently X-ray ``hot'' meaning that low-$Z$ metals can be fully ionised in the gas even  before receiving any irradiation by the X-ray corona. To account for this extra ``thermal ionization'', \citet{refbhb} performed self-consistent calculations of the reflection resulting from the illumination of the accretion disk by both a hard, powerlaw corona and thermal disk blackbody radiation. The authors compared the results of having the disk both in hydrostatic equilibrium and under the assumption of a constant density atmosphere, and found reasonably  good agreement between the two emergent spectra. \citet{refbhb} also confirmed in stellar mass black holes the result that had been previously found for AGN in that the spectrum from a constant density disk is slightly diluted (it has a lower  flux)  in comparison to that of a disk in hydrostatic equilibrium. Furthermore, the authors also noted a few small differences between the modes; namely a lower  effective ionization parameter in  the constant density model which resulted in a slightly stronger Fe\ka\ line and deeper iron K-edge. Nonetheless, the overall spectrum from the constant density approximation was shown to be in  good agreement with the result for an atmosphere in hydrostatic equilibrium. The two reflection grids resulting from the work of \citet{reflionx, refbhb}, will be used frequently throughout this work.

\subsection{General Relativistic Effects: Light bending}

Naively,  assuming  isotropic coronal emission, one would expect variations in the  reflection component to be directly correlated to variations in the observed power-law continuum.   That is, as the observed flux of the  X-ray corona increases, so should the amount of reprocessed emission. However, in a number of instances it has been found that this is not the case, with the reflection component at times behaving in an anticorrelated manner \citep[e.g.][]{Rossi2005j1650} or not varying at all despite large variations in the X-ray powerlaw continuum \citep[e.g.][]{Fabian02MCG, Fabianvaughan03, MiniuttiFabianGoyder2003,BallantyneVaughan2003,LarssonFabian2007}.

By virtue of its proximity to the black hole, the emission from the corona is naturally  affected by general relativistic (GR) effects. Some of the radiation from the corona which would otherwise escape is gravitationally focused  -- ``bent" --  towards  the accretion disk giving rise to enhanced reflection and selectively decreasing the X-ray continuum at infinity. A number of studies \citep[for instance][]{MartocchiaMatt1996, MartocchiaKaras2000LB, Miniu04, MiniuttiFabianMiller2004j1650, Niedzwiecki2008} have investigated  the effect of GR on a compact,  centrally concentrated X-ray corona close to a black hole\footnote{Observational  evidence for such compact X-ray corona has recently come from microlensing results where the size of the X-emitting region has been shown to be of the order of $\sim10$\rg\ \citep{ChartasKochanek2009quasar,DaiKochanek2010quasar, ChartasKochanek2012quasar, MorganHainline2012quasar}.}. The ``light-bending" model put forward by  \citet{Miniu04} predicts a number of semi-distinct regimes affecting the variability of the reflection component compared to the X-ray continuum:

\begin{description}
\item[Regime 1]  When the corona is very close to the black hole (a few gravitational radii \rg=$GM/c^2$), a large fraction of the radiation is bent onto the the accretion disk thus significantly reducing the amount of observed X-ray continuum and enhancing the reflection. A very steep emissivity profile is expected as the source is highly concentrated in the inner regions and the reflection is expected to be a steep function of the continuum in a quasi-linear manner. 

\item[Regime 2] When the central corona is slightly further from the black hole (at heights of $\sim 10$\rg), light bending causes the reflection component to vary significantly less than the  X-ray continuum. The amount of light bent towards the black hole decreases as the corona moves further from the black hole and the X-ray continuum increases.

\end{description}

 Finally, at heights  $\gg 20$\rg, light-bending becomes less important and the observed  continuum increases\footnote{ In the original paper of \citet{Miniu04}  a further regime -- Regime 3 -- was defined at large radii where the reflection and powerlaw flux appeared to be anti-correlated with one another. This was an artefact of  having a finite boundary for the disk extent of 100\rg, and instead the reflected flux should asymptotically become flat with respect to the continuum as a continuation of Regime~2 \citep[e.g.][]{Niedzwiecki2008}.}.    In this manner, the presence of gravitational light-bending has been invoked to explain the fact that Seyferts  \citep[e.g.][]{FabZog09, Fabian20121h0707} and XRBs \citep[e.g.][]{reismaxi}  at times appears to be ``reflection-dominated". Sources where the observed X-ray spectrum display a distinct lack (or comparably small amount) of hard, powerlaw-like continuum despite a large contribution of reflection are fully consistent with the first Regime detailed above.

Of course, the model presented above is idealised in that all characteristics of the observed variabilities are assumed to be a result of variation in the height of an isotropic  and compact corona with a fixed luminosity. Although intrinsic variation in the luminosity of the corona may well be present, it is unlikely that they could solely explain the behavior of the reflected emission described above. Indeed, the  clear presence of broad and skewed iron  emission lines in a growing number of sources ranging from stellar mass black holes \citep{miller07review,miller09spin, reisspin, Steiner2011, hiemstra1652} to AGNs \citep{tanaka1995,Nandra97, Nandra07, FabZog09,3783p1} (and also to lesser extent neutron stars \citep{cackett08, cackett10, DiSalvo2005170544, disalvo09, reisns}) strongly attest that general relativistic effects play an important role in producing the line profile, further  supporting the notion that the corona and the inner disk are both in the inner regions around a black hole.

\subsection{\j}

Amongst one of the first systems to provide observational evidence for the aforementioned  effect of gravitational light bending around a black hole and indeed the first around a stellar mass black hole was \j, which was discovered  by \rxte\ on 2001 September 5 as it went into outburst  \citep{Remillard2001}. Based on the spectrum obtained early in decay of the outburst by \xmm, and more importantly on the presence of a clearly broad iron emission line, \citet{miller02j1650} were able to not only infer that the object in \j\ was indeed a black hole, but also that it was close to maximally rotating with a dimensionless spin parameter $a$ $\approx0.998$. 

By decomposing the hard X-ray continuum from the reflection component in three \bepposax\ observations of \j, \citet{MiniuttiFabianMiller2004j1650}  were able to show that the  latter remained nearly constant despite a large change in the direct continuum, in a manner consistent with the predictions of light bending around a black hole.  Optical observations obtained after the system had returned to near quiescence  \citep{Orosz2004J1650} revealed a mass function $f(M) = 2.73\pm 0.56\msun$ with a most likely mass of $\sim4\msun$, and with it secured \j\ as genuine black hole  binary system. \citet{CorbelFender2004} reported on the radio and X-ray observations of \j\ during the outburst. The authors find a clear drop by nearly an order of magnitude in the radio flux at the transition from the hard intermediate state (referred to as the intermediate state in that work)  to the soft intermediate state (referred to as the steep power-law state in that work), and surprisingly they find residual radio emission during the often radio quiet disk-dominated soft state  which they attributed  to possible emission from previously ejected material interacting with the interstellar medium, rather than originating in the central source.

A follow up study of \rxte\ data by \citet{Rossi2005j1650} used the iron line as a proxy to the total reflection component and confirmed the plausibility of the light-bending scenario for the evolution of \j. Again using data obtained from the \rxte, \citet{Homan2003} reported on the discovery of a $\sim250$Hz QPO together with a number of less coherent variability peaks at lower frequencies. By studying the spectral and timing evolution during the first $\sim 80$ days of the outburst, the authors were able to define six  periods (I--VI; in this work referred to as P1--P6) having somewhat distinct   spectral and timing characteristics (see their Fig~1 and Table~1). A recent study involving  \j\  has discussed the similarities between X-ray binaries and AGNs \citep{Waltonreis2012} and argued that both \j\ and the active galaxy MCG--6-30-15 \citep{tanaka1995} must contain a rapidly rotating black hole, with the spin of \j\ having been formally constrained to $0.84\leq a \leq 0.98$.  A further body of work based on \rxte\ observations and a variety of empirical models for the hard X-ray continuum   \citep{YanWang2012} has concluded that the emission region, here referred  to as the corona, decreases by a factor of $\sim 23$ in size during the transition from the hard to the soft state.  

\subsection{This Work}

\begin{figure*}[!t]
\centering
{ \rotatebox{0}{
{\includegraphics[height=5.5cm]{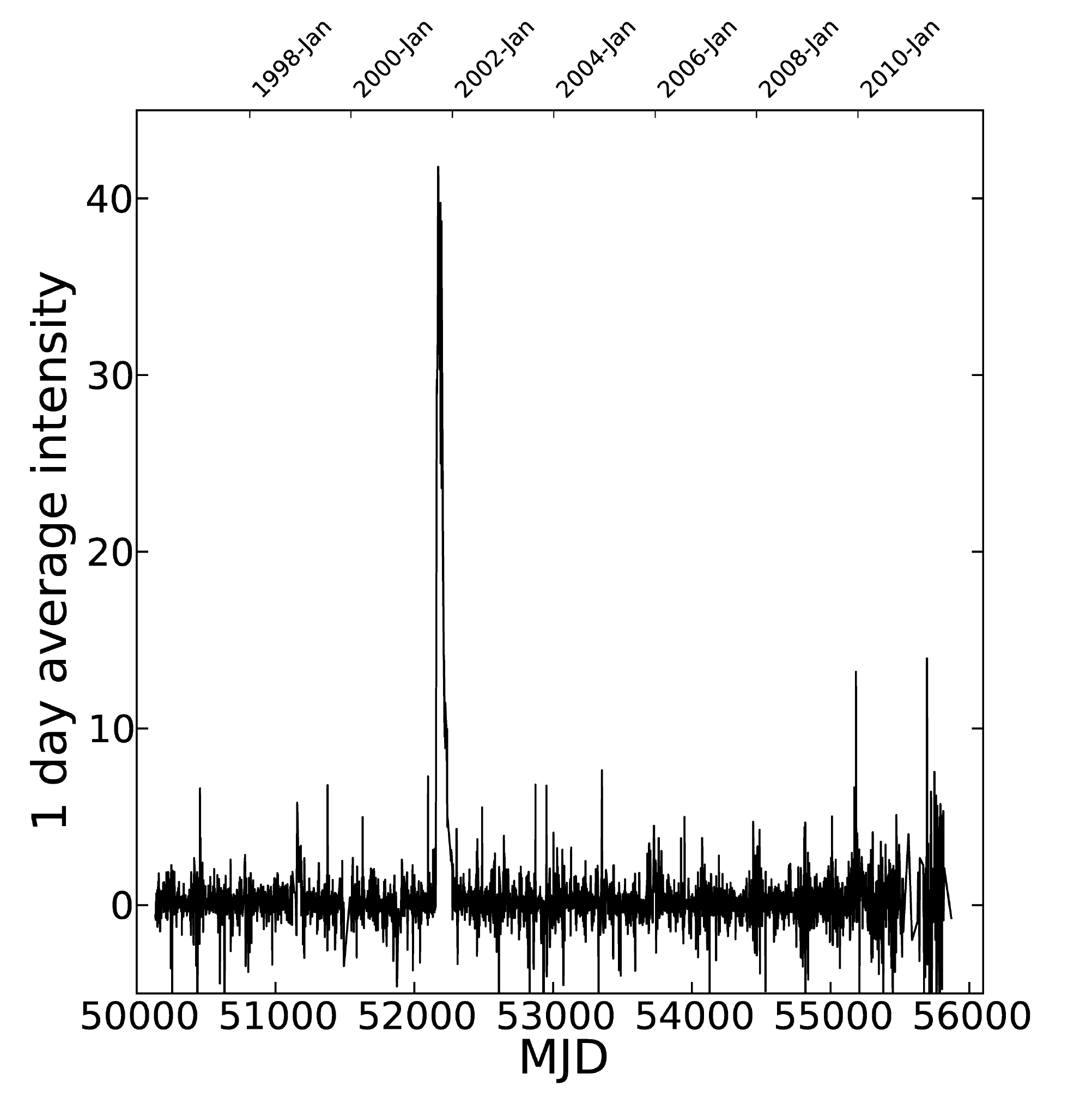}}}}
\hspace{-0.3cm}
{ \rotatebox{0}{
{\includegraphics[height=5.5cm]{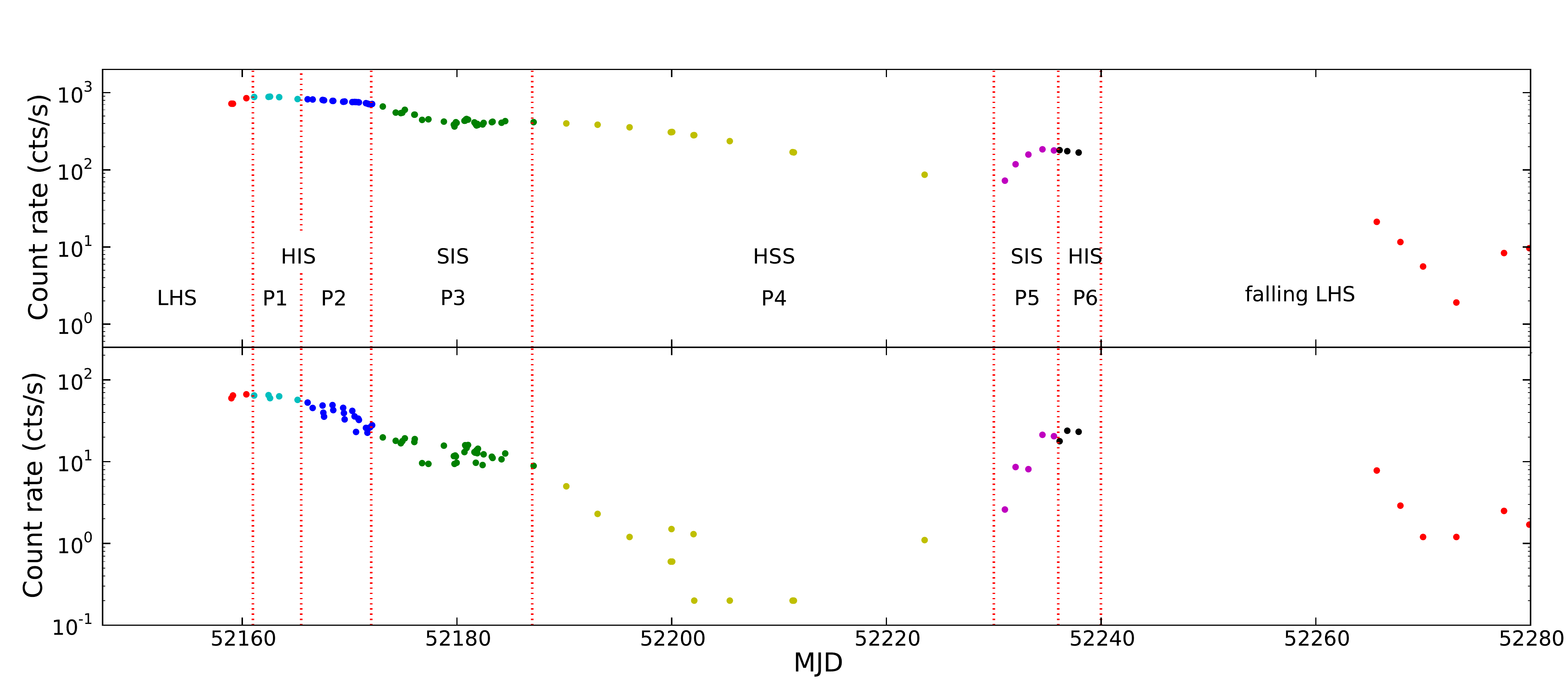}}}}

\vspace{-0.2cm}
\caption{\label{fig1} The 181 observations used were taken during the 2001/2002 outburst which is clearly seen in the All Sky Monitor (left). Since its discovery outburst, \j\ has remained in quiescence. (Right:) PCA-PCU2 (top) and HEXTE-A (bottom) count rate during the time encompassing the outburst. The different colors marks the distinct spectral states as defined by Homan et al.~(2003) based on the timing characteristics of the source. }  \vspace*{0.2cm}
\end{figure*}

Since its discovery, \j\ has become one of the best studied black hole systems. However, the energy spectra of this system have either been studied in a great degree of detail using high-quality, single snapshot observations with \xmm\ \citep[i.e.][]{miller02j1650, Waltonreis2012} or \bepposax\ \citep{MiniuttiFabianMiller2004j1650} or using mostly phenomenological and simple models in the study of long term evolutions with \rxte\ \citep[i.e.][]{Rossi2005j1650, Dunn2010, Dunn2010disk, YanWang2012}.  In this paper, we use the full \rxte\ archival data of the outburst to {investigate the evolution of the direct power-law continuum, reflection and thermal disk  components using, for the first time,  a fully self-consistent prescription for the reflection component}. 

In this manner, we combine the virtues of detailed analyses of single observations with the immense diagnostic power of multiple \rxte\ pointings. By being able to  decouple the total reflection component (Fe-\ka\ emission line together with all other reflection signatures) from the illuminating continuum, we find that the transition from the hard-intermediate state to the soft-intermediate state is accompanied by a sharp increase in the strength of the reflected emission in comparison to the direct continuum. We interpret this increase in the reflection fraction as a sudden collapse of the corona as the system approaches the thermal state, although we note that this may not be a unique interpretation.

This paper is structured as follows: \S2 briefly introduces the observations and details our various selection criteria. \S3 describes the base model and assumptions used throughout this work. The various results are presented in \S4 and in \S5 we present a qualitative picture of a possible physical scenario that combines all our findings.

\section{Observation and Data Reduction}
\label{observation}

We downloaded and analysed all 181 individual RXTE pointed observations of \j\ following its discovery.  This gave a total of 307.4\ks\  PCA exposure which were reprocessed with the latest \heasoftv\ tools.  As such, we followed all the well established, standard reduction guides\footnote{Found at \href{http://heasarc.nasa.gov/docs/xte/recipes/cook\_ book.html}{http://heasarc.nasa.gov/docs/xte/recipes/cook\_book.html}}. Given that it is the only PCU that is always on, as well as being the best calibrated of all units,  we used only data from PCU-2. We chose to use the standard offset  value of $<0.02$ as well as elevation above the Earth's limb $>10$\deg. Background spectra were produced using \pcabackest\ using filter files created using \xtefilt. The latest PCA-history file was also used and throughout this work we use PCA data between  3\kev\ (ignored channels  $\leq6$) and $25$\kev\ without additional systematic errors as we are mostly interested in relative changes and the impact on the errors of the various parameters as well as in the $\chisq$ distribution shown in Fig.~6  is minimal with the only change being a systematic shift to lower values.

For our analyses, we require both PCA and High Energy X-ray Timing Experiment (HEXTE) data (but see below). The HEXTE data were also reduced in the standard manner following the procedures outlined in the \rxte\ guide.  Background files were generated using \hxtback\ and the spectra were corrected for deadtime using \hxtdead. The appropriate response  and ancillary files were downloaded\footnote{From \href{ftp://legacy.gsfc.nasa.gov/xte/calib_data/hexte_files/DEFAULT/}{ftp://legacy.gsfc.nasa.gov/xte/calib\_data/hexte\_files/DEFAULT/}}. HEXTE data were fit between 25-150\kev.

\begin{figure}[!h]
\vspace{0.cm}

\hspace{-0.9cm}
{ \rotatebox{0}{
{\includegraphics[width=9.5cm]{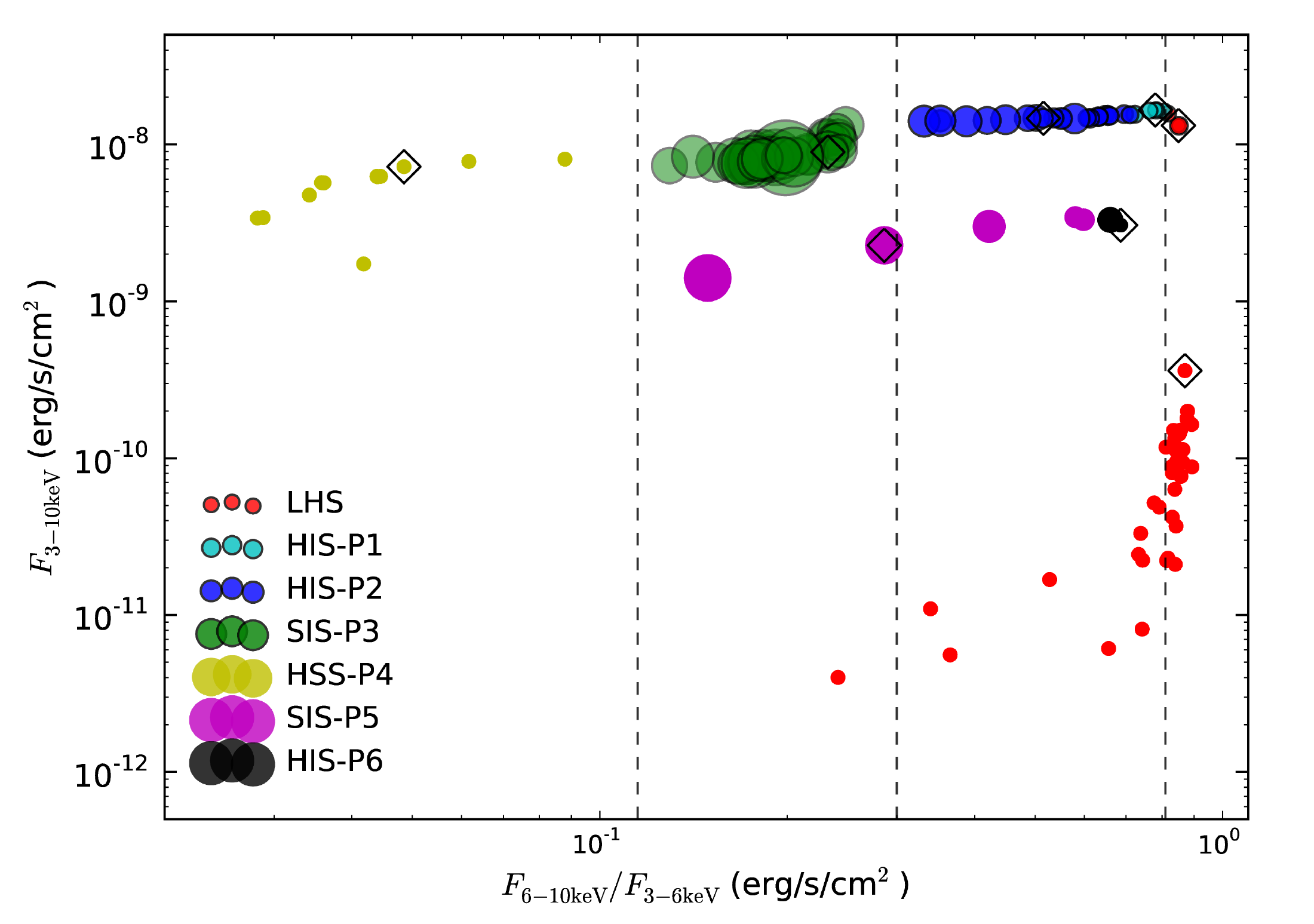}}}}
\label{fig2}
\vspace{-0.4cm}
\caption{Hardness intensity diagram made using the 116 observations that remain after imposing a cut where both HEXTE Cluster A and B have source-background counts greater than 0. The vertical dashed lines marks approximate transitions between LHS - HIS- SIS -HSS.  The large diamond symbols mark the position in the HID diagram for the representative spectra shown in Fig.~4. The color code is identical to Fig.~1 (right) and will remain the same throughout  this work. We have scaled the size of the symbols for each observation to the value of the reflection fraction $R$, as determined in \S~4.1 and shown in Fig.~3. The size-legends from top-to-bottom are: $R=0.5, 0.75, 1, 2, 3, 4, 5$. The falling branch of the LHS and the HSS have sizes corresponding to $R=0.5$ as $R$ is poorly constrained in these states.  }  
\end{figure}

Figure~1  (left) shows the 1-day averaged long term light curve as seen from the \rxte\ All Sky Monitor (ASM) where the 2001 outburst is clearly visible. The panels on the right shows the PCA-PCU2 count rate (top) and HEXTE-A (bottom) during the time roughly encompassing the outburst.  The colours  highlight the various states during the outburst and are directly mapped into the hardness intensity diagram (HID) shown in Fig.~ 2. In short, the first few observations caught the source in a rising LHS which evolved to the HIS approximately 3 days later, where it remained for $\sim15$ days until the clear change to the SIS. We have further divided the HIS into early Period 1 and late Period 2, similar to the division made by Homan et al.~(2003) based on the timing characteristics displayed. The source then remains in the SIS (P3) for approximately 12 days before it makes the typical excursion to the disk dominated HSS (P4) where  we see a clear drop in the HEXTE count rate in Fig.~1. After $\sim40$ days, the hard flux in \j\ sharply increases (P5 and P6) as the system returns to LHS and eventually goes back into quiescence where it remains up to the present time.

The work presented in this paper is fully interpreted within the framework of reflection models \citep[see e.g.][and \S1.1]{Fabianross2010}. As such we are mainly interested in being able to determine the contribution of the reflected emission to the overall spectra. As mentioned in the introduction, reflection is not limited to the iron line profile and in order to obtain the best handle on the full reflection component we restricted the analysis that follows by requiring that {\it both} the HEXTE-A and the HEXTE-B units have background-subtracted count rates that are greater than zero.  This results in 116 PCA pointing observations totalling 217.9\ks. In doing so,  we are effectively reducing our sensitivity to observations in very steep or disk dominated states, as well as to faint LHS, and most of the results presented here concern the intermediate states, which are more luminous at high energies (see Fig.~1; right). In order to investigate the effect of the HEXTE data on the results presented in this paper, we  have also repeated all the fits described below using only the PCA data and confirmed that, although the results obtained do not strictly  depend on the inclusion of the HEXTE data, owing to its relatively small statistical weight, the high energy data still provide a useful additional test of the continuum model  (see e.g. Fig~4).

In the following section we detail the spectral fits to the 116 observations. All our work makes use of the X-ray fitting package \xspecv\ \citep{xspec}.  Where uncertainties on models parameters are quoted, this refers to the 90 per cent confidence limit for the parameter of interest.

\label{Analyses}

\section{The model}
Previous attempts to characterise the evolution of \j\ as observed with \rxte\ have relied on a purely phenomenological interpretation of the reflection continuum and features\footnote{Contrast this phenomenological approach to the {\it reflection} component with the systematic testing of Comptonisation models for the {\it power-law} (corona) component  by \citet{NowakWilmsDove2002gx} on \rxte\ data of GX~339-4. Ideally one would strive to  combine physical models for both the power-law and reflection (and of course the accretion disk which is known to not be a simple multicolour disk due to various relativistic effects) however this quest for fully relativistic disk and reflection together with physical prescription for the Comptonisation continuum  is proving highly complicated even for single, dedicated efforts at snapshot observations and is  beyond the scope of this paper.}  and often employed a single \gaussian\ emission line with centroid energy fixed at 6.4\kev\ as expected from neutral Fe-\ka\  \citep[i.e.][]{Dunn2010disk},  a combination of a similar \gaussian\ together with a broad smeared edge \citep[\smedge,][]{smedge} component (i.e.~Yan \& Wang 2012), or in the more physically appropriate application, a relativistic emission line \citep[\laor,][]{laor} together with \smedge\ (Rossi et al.~2005). However, even in the latter example, the model was unphysical as the combination of \smedge\ with an emission line does not keep consistency between the depth of the edge -- the number of absorbed photons -- and the strength of the line. 

 In the study of  \citet{Dunn2010disk}, where the authors were primarily interested in the behavior of the thermal accretion disk, such a simplification is easily justified as small deviations at high energies are unlikely to have significant effects at low energies. Rossi et al. alluded at the importance of using reflection models but unfortunately were swayed away due to the highly time-consuming and computer-intensive task that this would present nearly 10 years ago.   

The importance of reflection from accretion disks can be directly measured by the sheer number of theoretical  works that have been devoted to fully characterising its behavior \citep[i.e.][]{LightmanWhite1988, George91, Matt1991,  rossfabian1993, Zycki1994, NayakshinKazanasKallman2000, cdid2001, reflionx,  refbhb, GarciaKallman2010, GarciaKallman2011}. A widely used reflection code is  that of \citet[][\rm{\small REFLIONX}]{reflionx}, which provides a self-consistent treatment of the dominant atomic processes around a black hole and, given an input power-law continuum, outputs a reflection spectrum where both the ``Compton-hump", emission and absorption features are {\it  all physically linked}.

The reflection spectrum is calculated in the local frame, therefore we employ the convolution model \kdblurf\ (\citet{laor}; vastly optimised by Jeremy~S. Sanders and employed in \citealt{fabian2012cyg} and \citealt{reismaxi}) to model the relativistic effects in the spectra. Despite the existence of newer relativistic models that includes the black hole spin (as opposed to the extent of the inner radius) as a variable parameter \citep[i.e.][]{BeckwithDone2004, kyrline,kerrconv,  relconv}  the advantage of   using \kdblurf\ on such a vast data set is  it greatly expedites the fits and since we are not overly interested in absolute values, the small improvements of the newer models at the 10\% level \citep{BeckwithDone2004} does not justify their~use.

Each of the 116 remaining spectra were thus fit with a base model described in \xspec\ as\footnote{We have also repeated this experiment with the latest reflection model by \citet{refbhb} (\refbhb; see also \citet{reisgx} for a description of its use) which incorporates a black body component and found the results to be consistent in both case. We choose to carry on with the $\reflionx + \diskbb$ combination for a number of reasons; to start, this combination  can be easily reproduced (\refbhb\ is not yet public) and the output parameters are somewhat standardised. e.g. \reflionx\ has $\xi$ as a free parameter whereas \refbhb\ has the more obscure combination of hydrogen number density $n_{\rm H}$ and the relative strength of the illuminating flux over the  thermal emission at the surface of the disk.} \\

\noindent{$\phabs\times(\po+ \diskbb+ \kdblurf\otimes\reflionx$)}. \\  

\noindent{To avoid unnecessary CPU cost, all spectra were fitted adopting the same initialising values for the model parameters. A similar approach of using a base model with similar starting parameters was taken by \citet{WilmsNowak2006cyg} -- amongst others -- in the study of Cygnus~X-1.}

Since the low-energy cutoff in the PCA of $\sim3$\kev\ is not sufficient to constrain the neutral hydrogen column density, throughout this work we have frozen the parameter in the \phabs\ model  to  $7\times 10^{21} ~{\rm atom}\pcmsq$ in \xspec\ as the neutral column density is likely best modelled as being constant with time \citep*{nhpaper}\footnote{We used the standard BCMC cross-sections \citep{balucinska} and ANGR abundances \citep{abundances}.}. The key parameters in the \reflionx\ model are the  spectral shape of the illuminating continuum which is set to be the same as  that of the direct power-law and the ionization parameter of the accretion disk  (as in Equation~1). The iron abundance was set to the solar value and the redshift was set to zero.

In order to make the work more tractable, the emissivity profile in the blurring component was initially restricted to  the form of a single power-law such that $\varepsilon (r) \propto r^{-q}$, and  following the most recent work on the best \xmm\ and \bepposax\ data of \j\ (Walton et al.~2012) we have frozen the inclination of the accretion disk  to 65\deg\ and note that this parameter is highly constrained based on optical light curve to be greater than $50\pm 3$\deg\ and $<80$\deg\ (Oroz et al.~2004). We also check whether the results presented below change if we freeze the inclination at this lower limit (50\deg) and we confirm that they do not.  As is standard with such fits, we have frozen the outer radius to the maximum value in the grid of 400\rg\ and the inner accretion disk is free to vary.

A phenomenological combination of a \laor\ emission line on top of a reflection model such as \pexriv\ 
\citep{pexrav} -- which does not include the iron-\ka\ emission line -- often provides an equally good fit to X-ray spectra of black 
holes. However, \pexriv\  does not Compton broaden its absorption edge nor does it provide a physical coupling between itself and 
the extra \laor\ component which can result in parameters severely lacking in physical consistency. This is exemplified in 
\citet{reismaxi} where we showed for the stellar-mass black hole candidate MAXI~J1836-194 that the reflection component  could be equally well modelled with a combination of \laor~+~\pexriv; \laor~+~\kdblur$\times$\pexriv; or \kdblur$\times$\reflionx.  The first combination unphysically required the iron line to be coming from far within 6\rg\ in the strong gravity regime, yet all other reflection features -- under the formalisation of the model -- appeared exempt from such effects. After properly correcting  for relativistic effects in \pexriv, the ionization parameter $\xi$ of the second model  was nearly two orders of magnitude  higher than the previous combination as the Fe absorption edge in \pexriv\ was no longer trying to fit the blue wing of the iron line. This change was also accompanied by an artificial hardening of the powerlaw index and a decrease in the equivalent width of the \laor\ component from $\sim 270$\ev\ to $\sim180$\ev.  Correctly blurring the originally sharp absorption edge caused it to become smooth and more symmetric. Due to the decoupling between the \laor\ line and \pexriv, this smooth edge traded off with  the \laor\ line component thus decreasing the equivalent width of the latter. As such, we stress the  importance of the imposed physical consistency in the  emission line and corresponding absorption edge strength  
afforded by \reflionx\ and strictly enforce this in the  work that follows.

\begin{figure*}[!t]

\begin{center}
{\vspace{.8cm}\hspace*{-0.8cm}\rotatebox{0}{{\includegraphics[height=20.5cm]{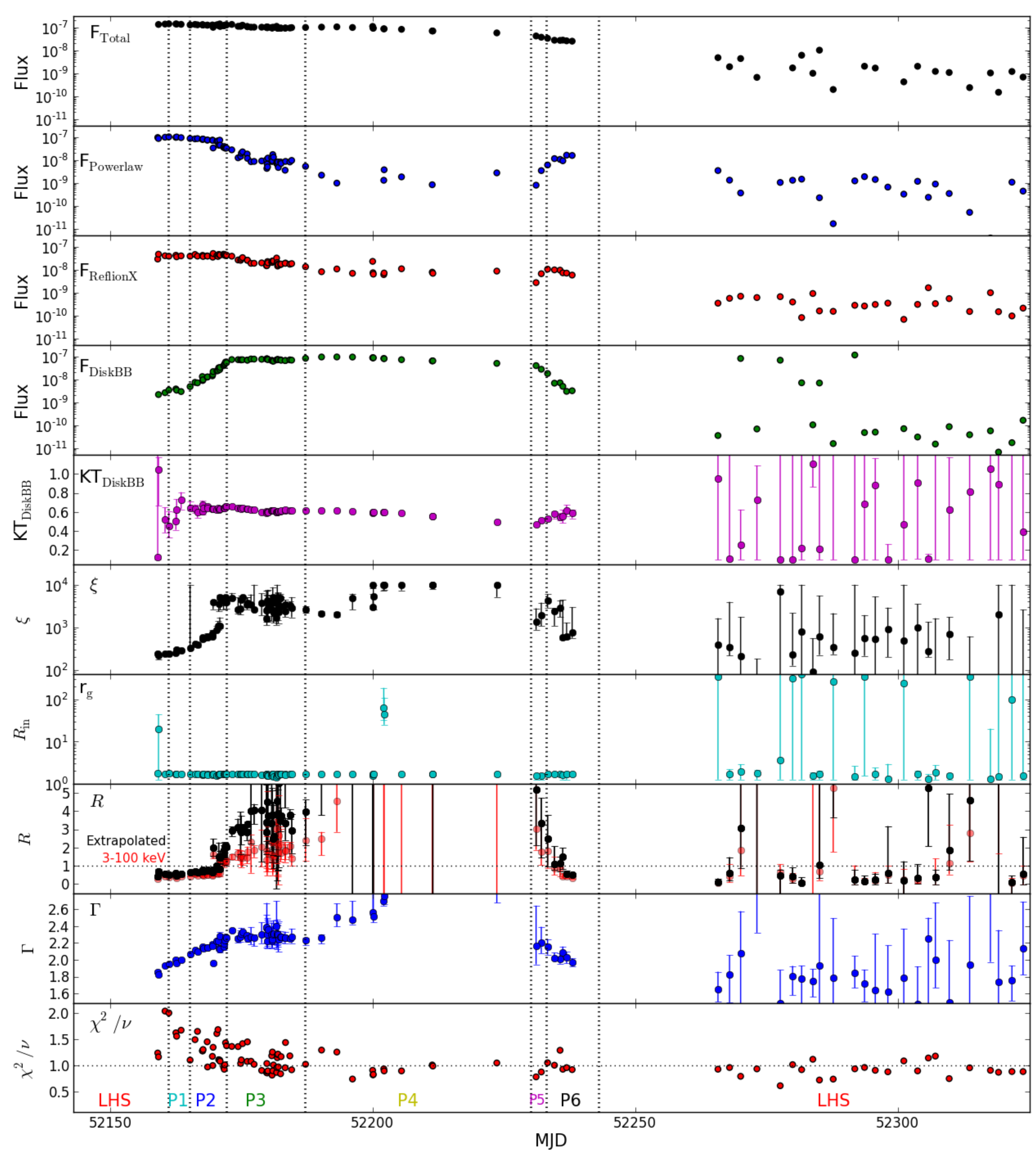} }}}
\caption{Top four panels: Extrapolated 0.1-1000\kev\ flux evolution for the total, powerlaw, reflection and disk components.   The errors in the fluxes were assessed using \cflux\  but are omitted here for clarity (but see Fig.~7 for an indication of the magnitude of the errors in each state). These errors were propagated in the derivation of the reflection fraction $R$ (eighth panel from the top). We also show in the eighth panel ($R$) the ratio of the 3--100\kev\ reflection to powerlaw fluxes. A jump is in this ratio seen at the same position as that for the extrapolated fluxes.  The bottom panel shows the reduced-$\chi^2$ for 114 D.o.F.  The vertical dashed lines show the transition from the rising LHS, P1--P6 and back into the falling LHS.  }  \vspace*{0.5cm}  
\label{fig3}
\end{center}
\end{figure*}

\vspace{0.cm}
\section{Results and discussion}

\subsection{Evolution of the Outburst}
\label{Light bending}

\begin{figure*}[]
\centering

\hspace*{0.cm}
{\hspace{-0.0cm}\rotatebox{0}{{\includegraphics[width=5.5cm]{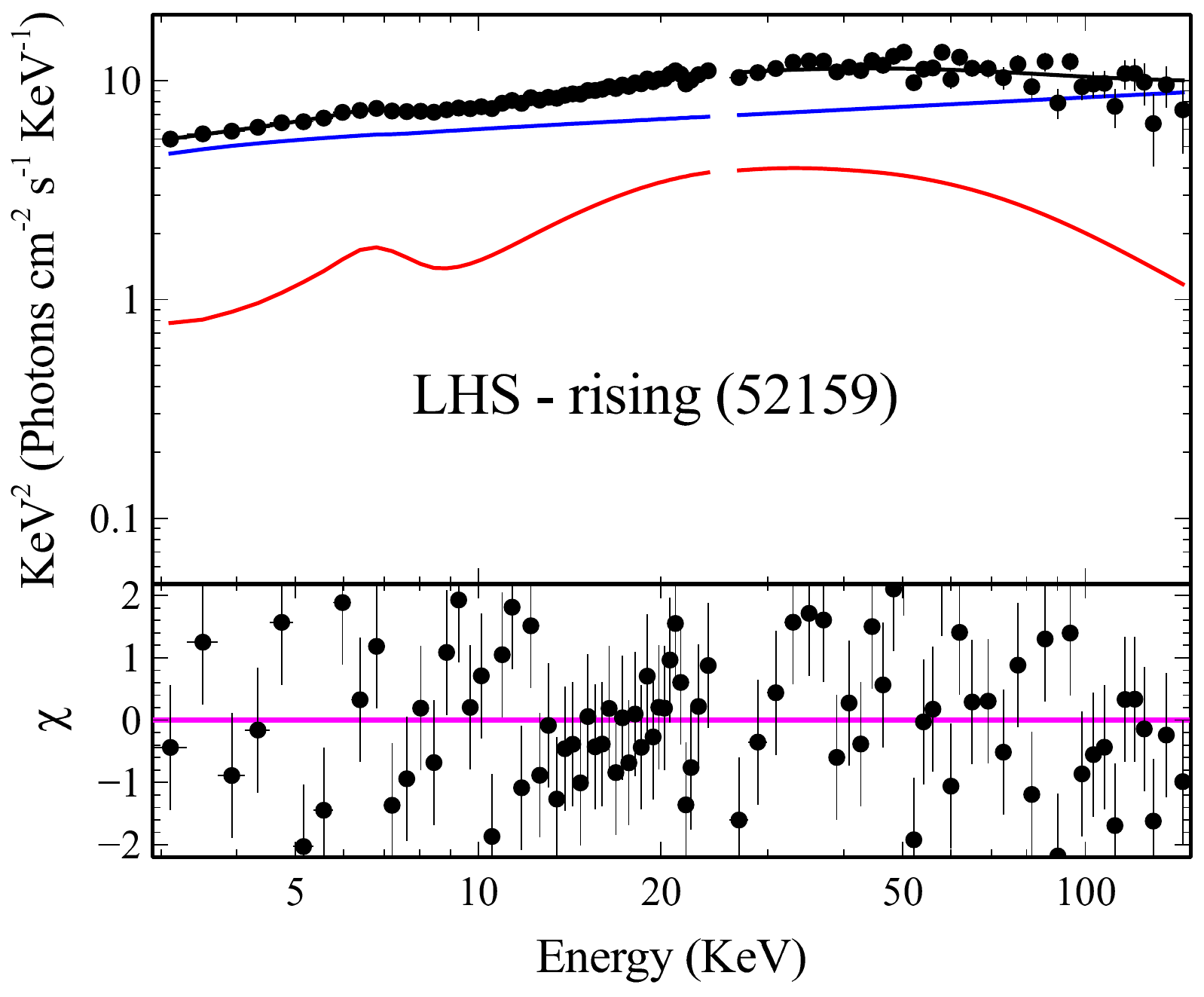} }}}
{\hspace{0.cm} \rotatebox{0}{{\includegraphics[width= 5.5cm]{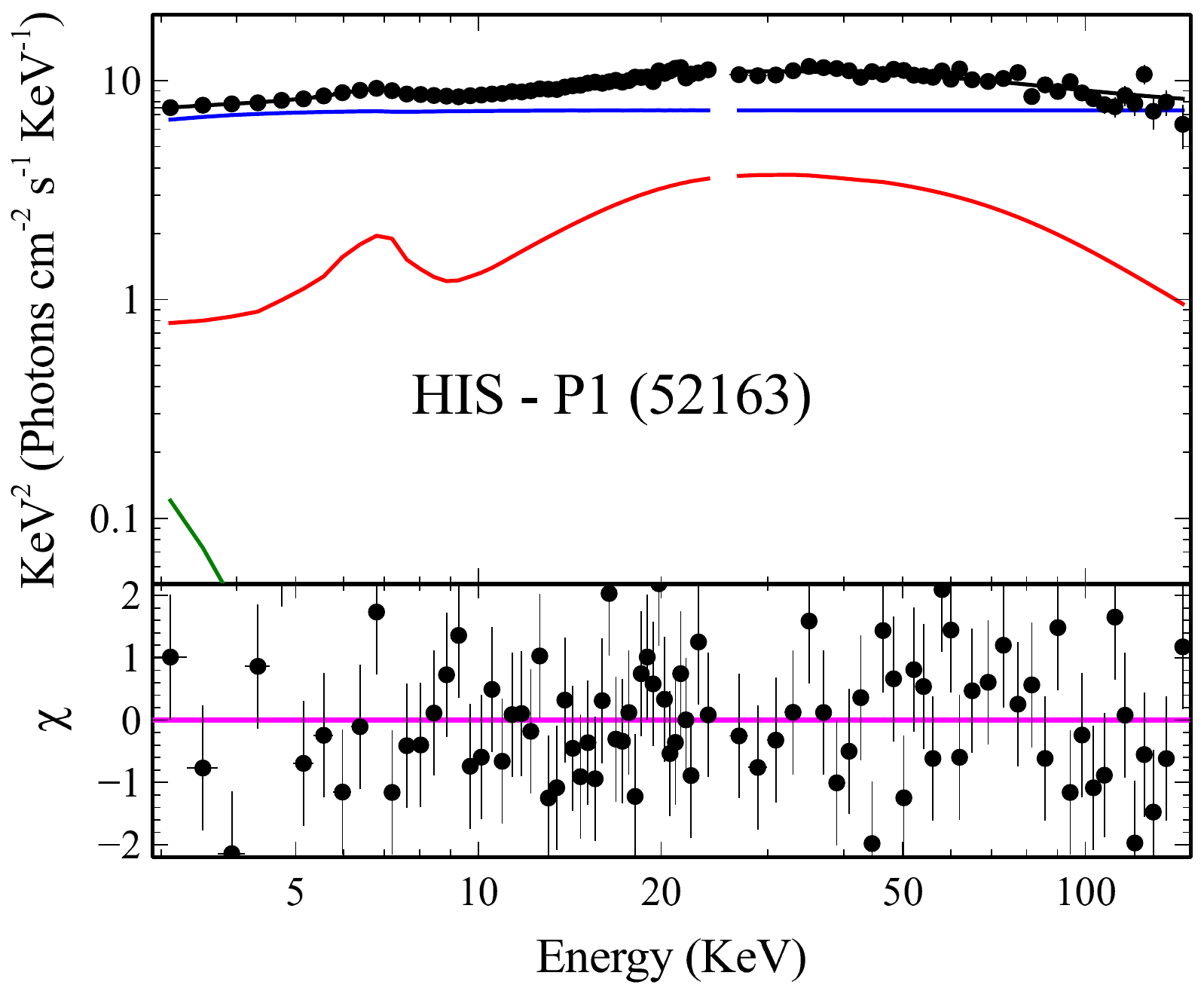} }}}
{\hspace*{0.cm} \rotatebox{0}{{\includegraphics[width= 5.5cm]{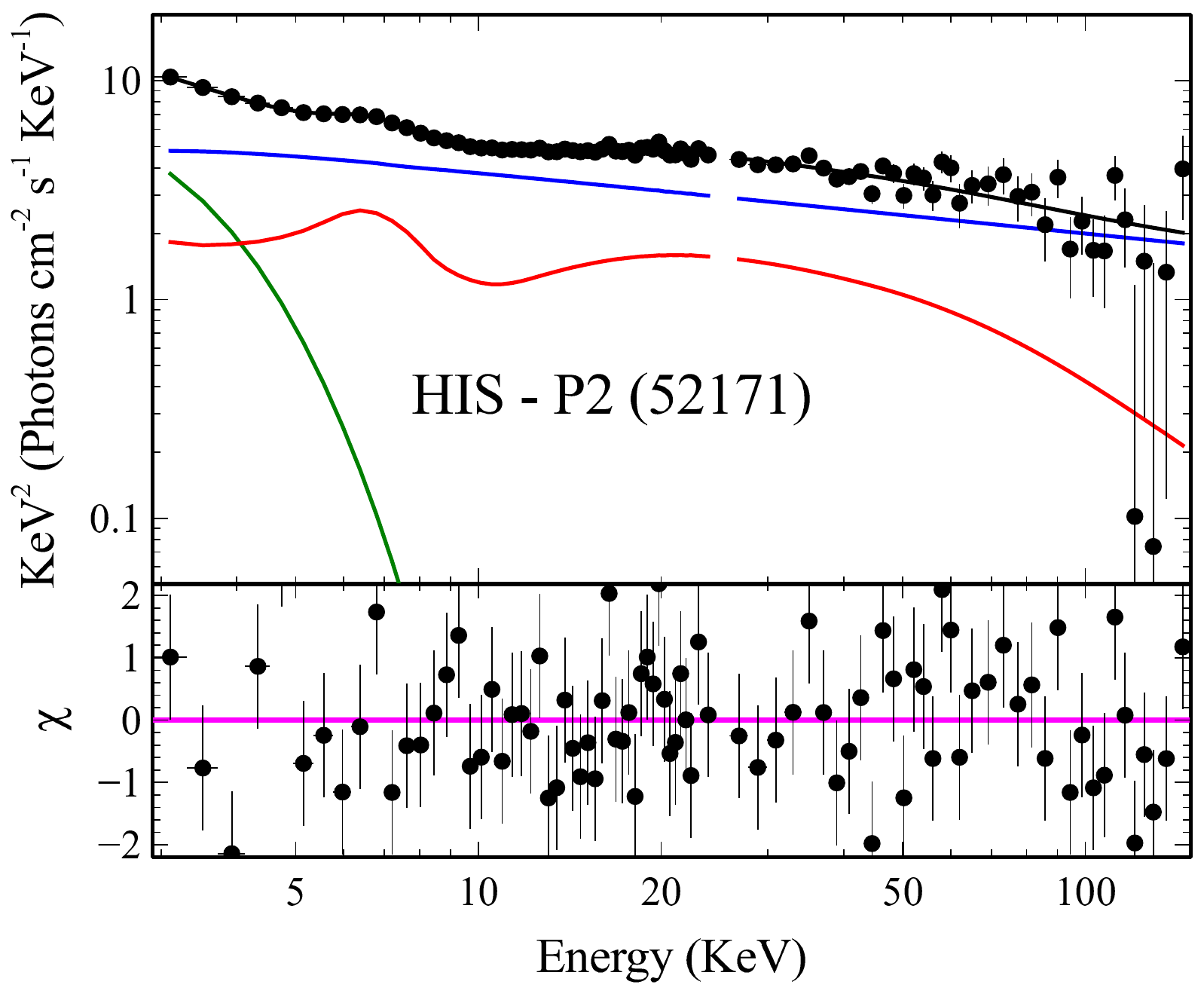} }}}
{\hspace{-0.0cm}\rotatebox{0}{{\includegraphics[width=5.5cm]{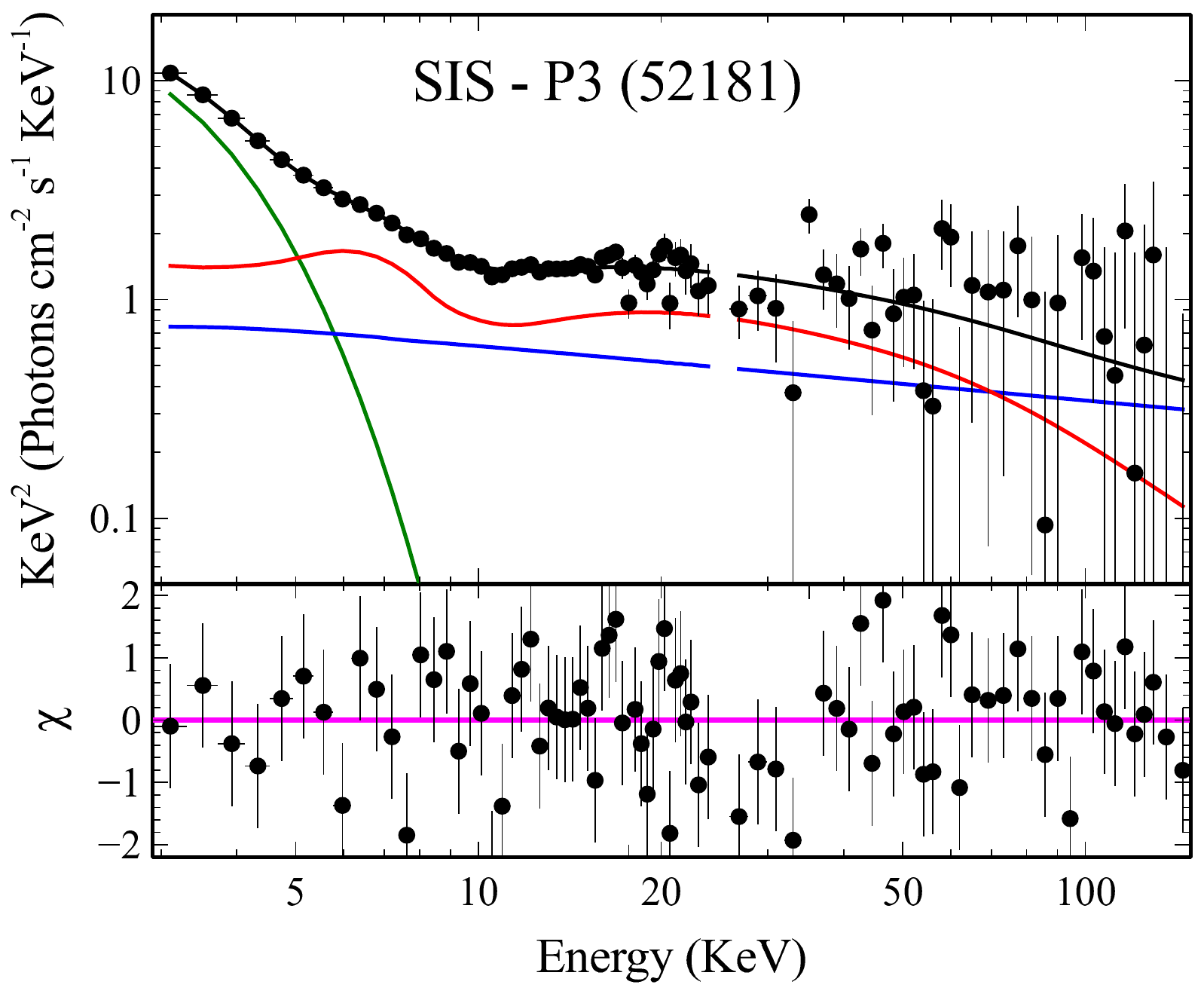} }}}
{\hspace{0.cm} \rotatebox{0}{{\includegraphics[width= 5.5cm]{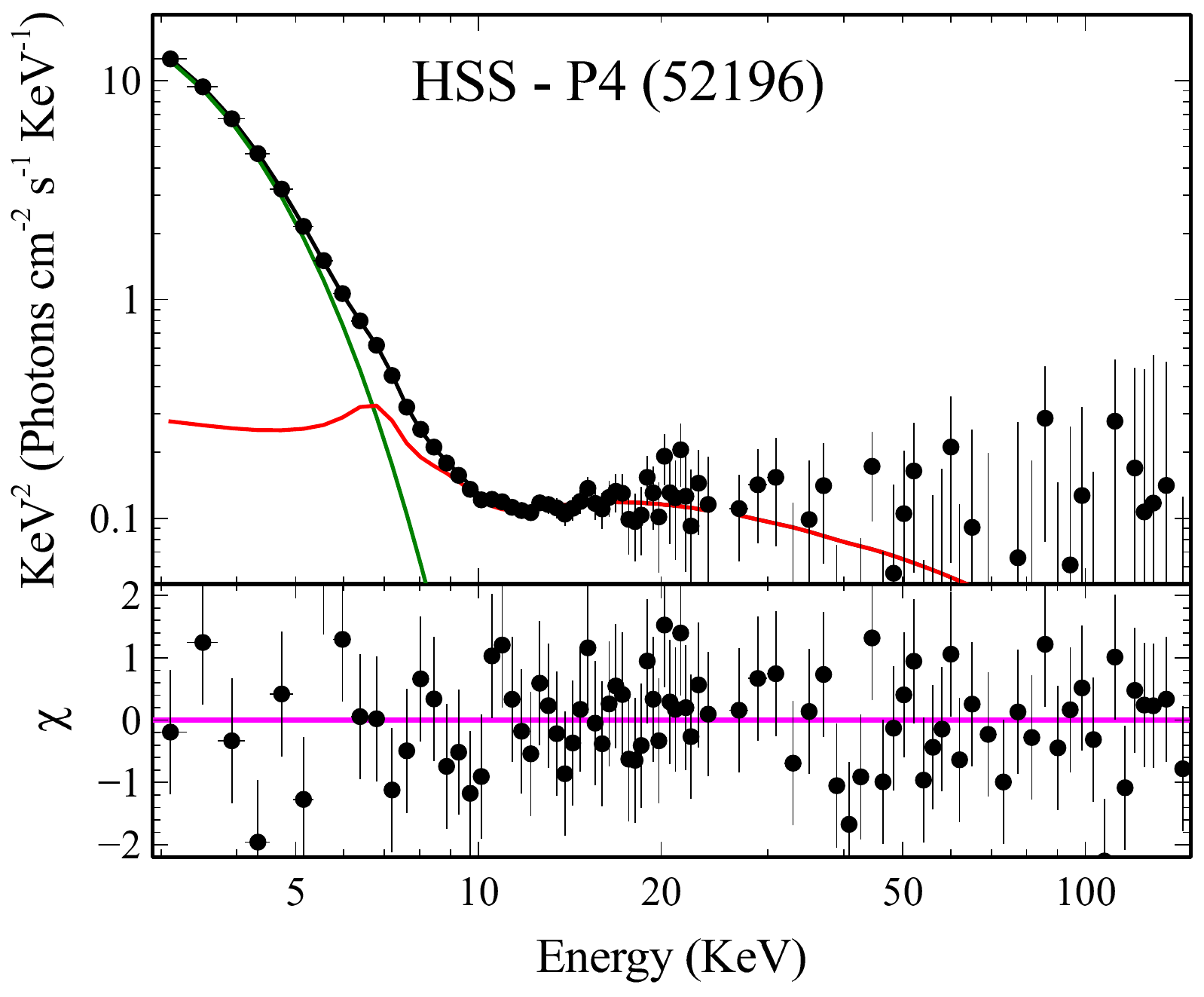} }}}
{\hspace{0.cm} \rotatebox{0}{{\includegraphics[width= 5.5cm]{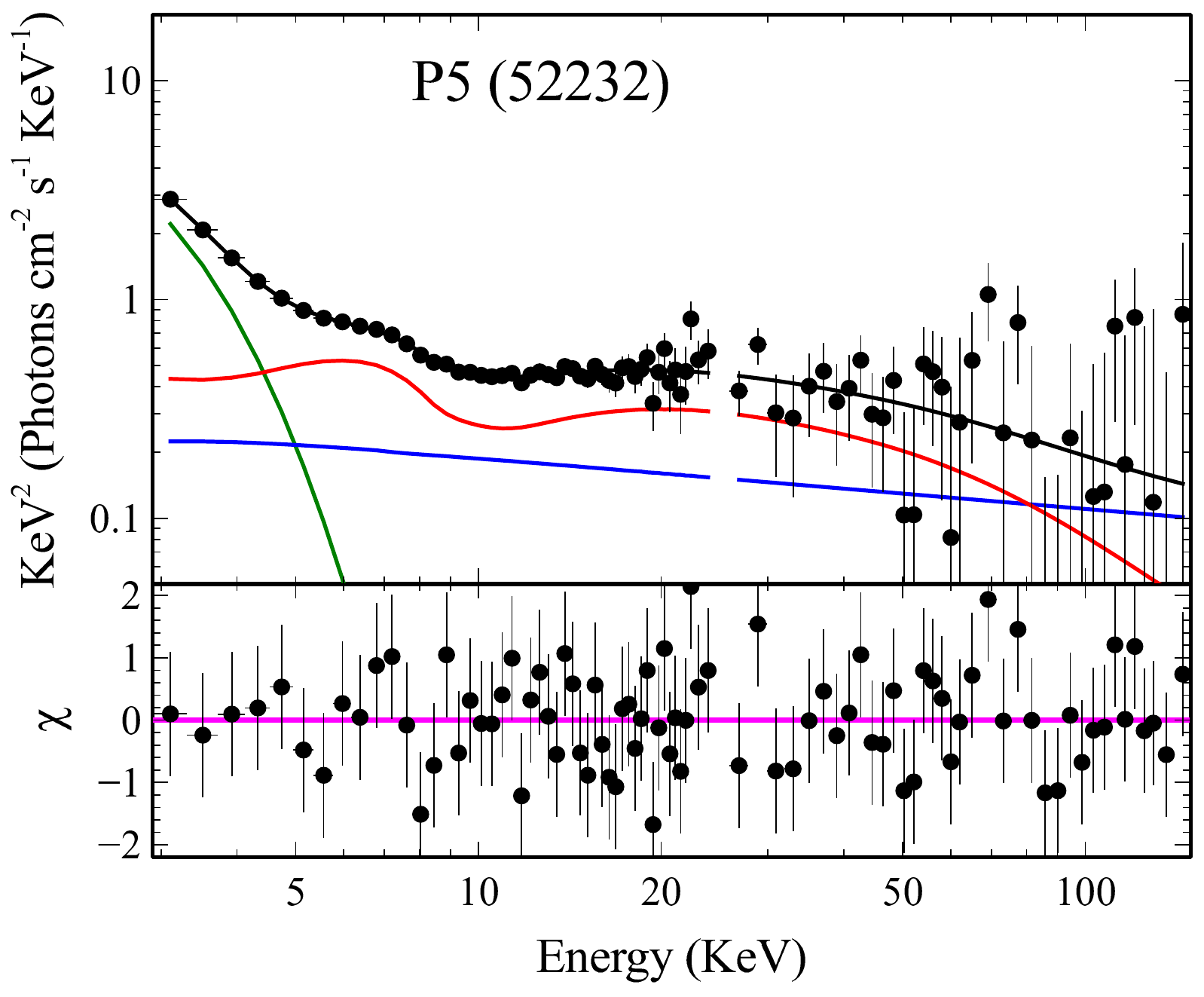} }}}
{\hspace*{0.cm} \rotatebox{0}{{\includegraphics[width= 5.5cm]{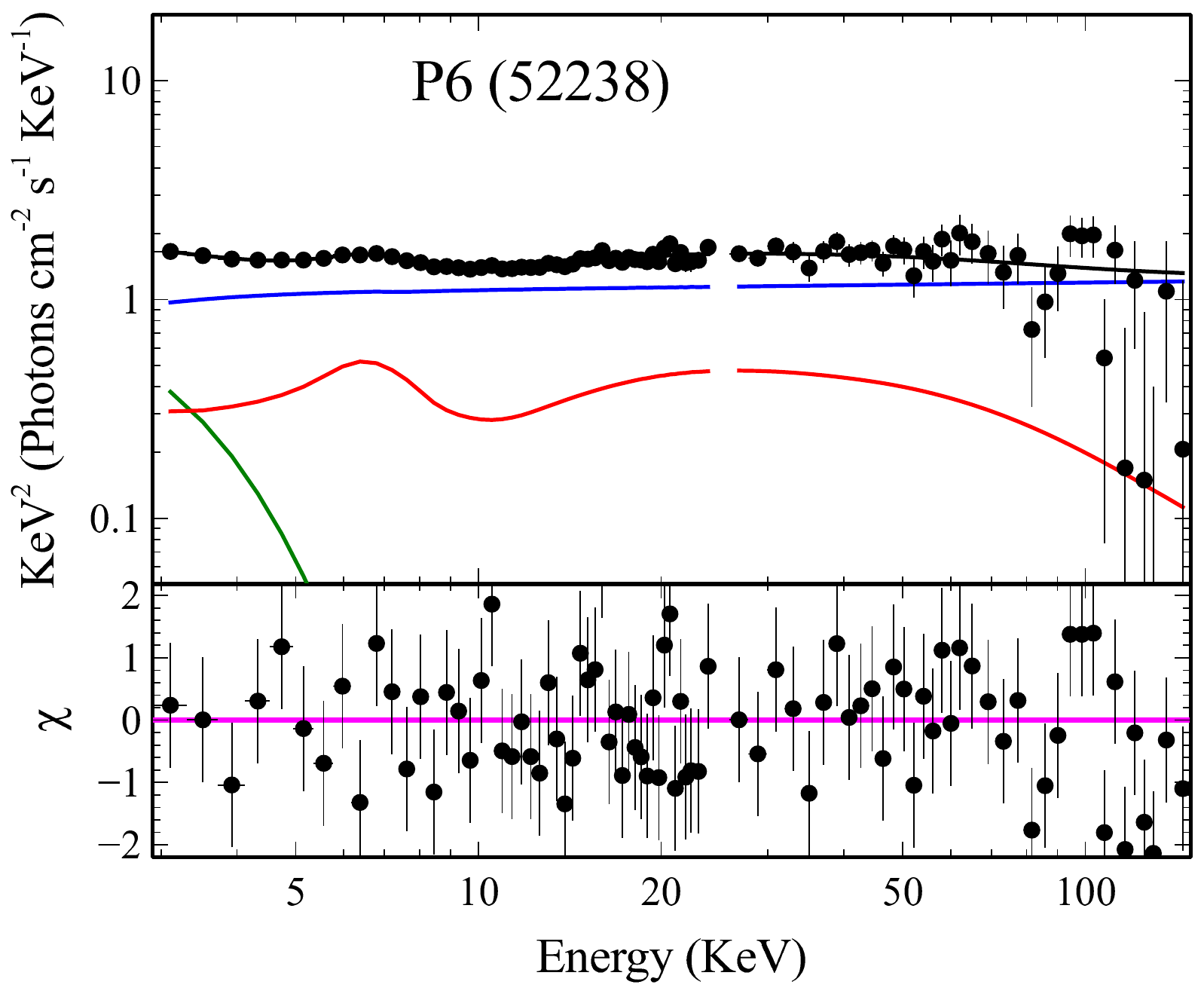} }}}
{\hspace{-0.0cm}\rotatebox{0}{{\includegraphics[width=5.5cm]{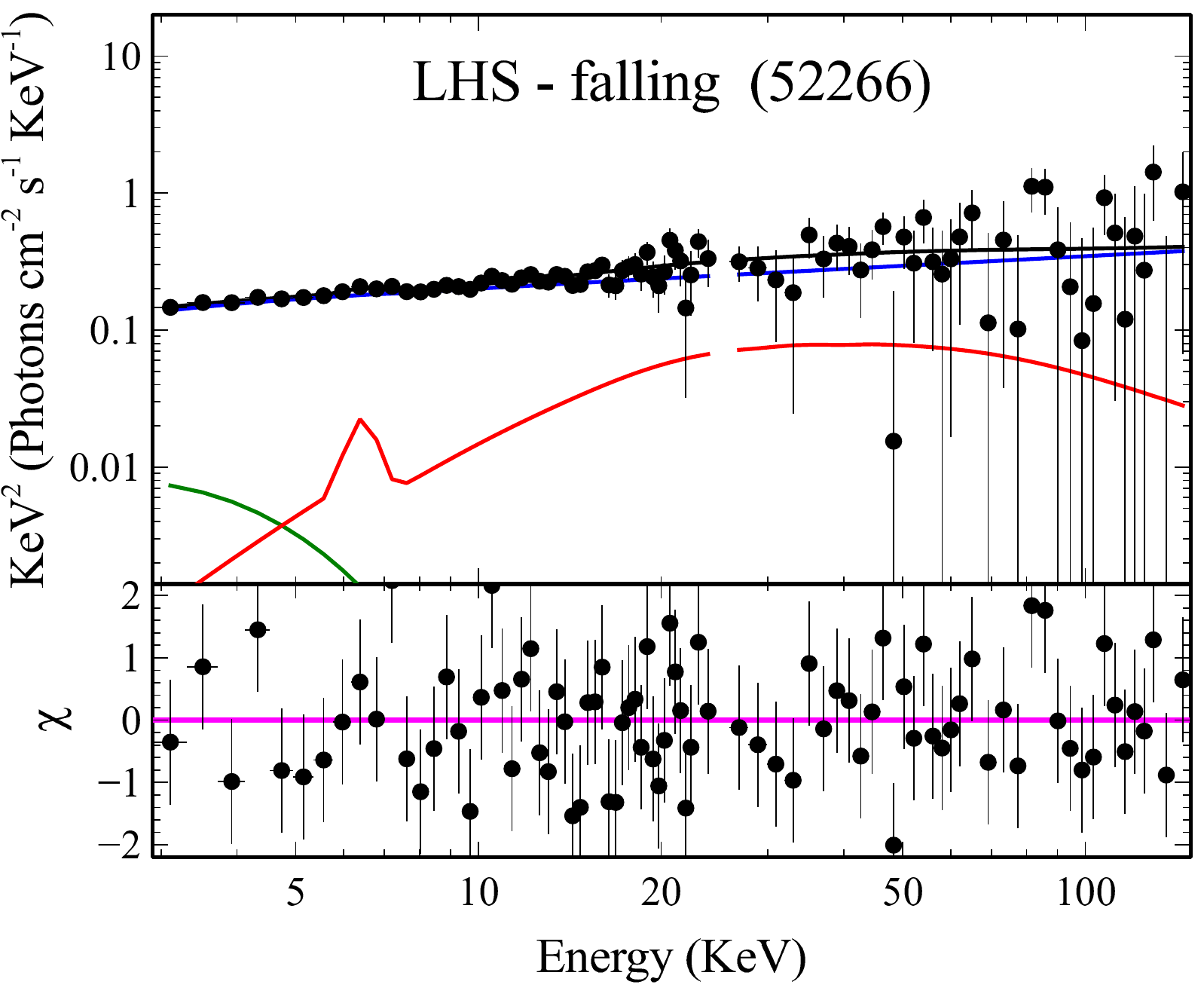} }}}

\caption{Unfolded spectra (top) and residuals (bottom) for the eight representative states shown in Figs.~1,2.  The reflection, power-law, and disk component are shown in red, blue and green respectively. The total model is shown in black.  All vertical scales are the same except for that of the falling LHS (Bottom right). The approximate MJD of each observation are shown in brackets. }  \vspace*{0.3cm}
\label{fig4}
\end{figure*}

Figure~3 shows the evolution of all the parameters of interest in this work\footnote{We refer the reader to the work of \citet{Dunn2010disk} for the evolution of the disk properties during the outburst. As mentioned above, despite the fact that the authors did not correctly account for reflection in their work, this plays a minor role at the energies of interest in regards to the disk properties and as such that work is still a valid and important reference for the evolution of accretion disks.}, and Fig.~4 shows the best-fit to eight representative spectra roughly covering the eight periods highlighted in Figs.~1,2 and described in detail in Homan et al.~(2003). The spectra used for illustration are shown in Fig.~2 with diamonds.  The top four panels in Fig.~3 show the evolution of  the extrapolated 0.1-1000\kev\ fluxes for the total, \po, \reflionx\ and \diskbb\ components, from top to bottom respectively. All fluxes were obtained using  \cflux\ in \xspec.  The vertical dotted lines running through all the panels highlight the eight periods shown in Figs.~1,2. We see a clear increase in the disk flux during the first $\sim30$ days followed by a clear flattening as it moves into the HSS. It is also visually apparent that the reflection flux varies relatively less than the power-law continuum. This will be investigated further in what follows. 

The following  two panels show the evolution of the disk temperature and  ionization parameter, $\xi$. Early in the outburst, the disk was relatively cold ($\lesssim0.5$\kev) and only moderately ionised with $\xi\approx200\ergcmps$. As the system moves into the HSS, the ionization increases smoothly until  $\sim2$~days before the transition to the SIS when $\xi$ sharply increases to $\xi\approx3000\ergcmps$ and remains at that level through the transition up to the end of the SIS. The disk temperature, on the other hand, appears to reach a relativity  stable value of $\sim0.6-0.65$\kev\ early in P2, approximately half way through the HIS. As the system progress into the HSS, the disks becomes  fully ionised with $\xi$ reaching the maximum allowed value in the model (log~$\xi=4$), before coming back down to the low hundreds towards the end of the outburst. 

The reflection fraction $R$ shown in the third-from-bottom panel of Fig.~3 is here defined as a measure of the ratio between the (observed) continuum power-law  to the reflection flux emitted by the disk. Since a fraction of the  power-law illuminating the accretion disk is down-scattered as it is reprocessed in the disk, the reflection fraction in \reflionx\ is calculated by dividing the  extrapolated  (1\ev - 1000\kev) \reflionx\ flux by the   0.1-1000\kev\ power-law flux.  At the start of the outburst, through to the end of the HIS, $R$ increases smoothly between   $\approx 0.6-1$. At the transition between the HIS to the SIS, $R$ displays a sharp increase to $\approx  4$ where it remains until the beginning of the disk dominated HSS where the power-law effectively disappears. We also show in this panel the ratio of the 3-100\kev\ reflection to powerlaw flux. The behavior described above is still qualitatively the same and we still see a clear jump in this ratio at the transition from the HIS to the SIS. However, when limited to the 3-100\kev\ flux, this ratio is systematically a factor of $\approx1.5$ less than the extrapolated ratio;  a direct  result of not accounting for the extra down-scattered flux at low energies.

As a further test of both the qualitative (clear jump in \textit{R} between HIS and SIS) as well as quantitative (change from $R\approx 0.6$ to $R\approx 4$ between P1 and P3)  behaviors found here for the reflection fraction, we temporally replace the \reflionx\ model with a combination of \laor\ plus \pexriv\ and employ this model to the observations highlighted in Figs. 2 and 4 for P1 and P3. In using this model, we have blurred the \pexriv\ component with the same parameters as the \laor\ line profile
 Figure~5 shows the confidence range for the reflection fraction (a free parameter of \pexriv)  for these two representative spectra. In agreement with our previous results, we see that in the HIS the reflection fraction is constrained to  $R=0.58^{+0.08}_{-0.11}$ and in the SIS it is $R>2.7$ at the 90~per~cent level of confidence ($\Delta\chi^2 = 2.71$).

\begin{figure}[]
\centering
\vspace{0cm}
{\hspace*{-0.2cm}\rotatebox{0}{{\includegraphics[width=9cm]{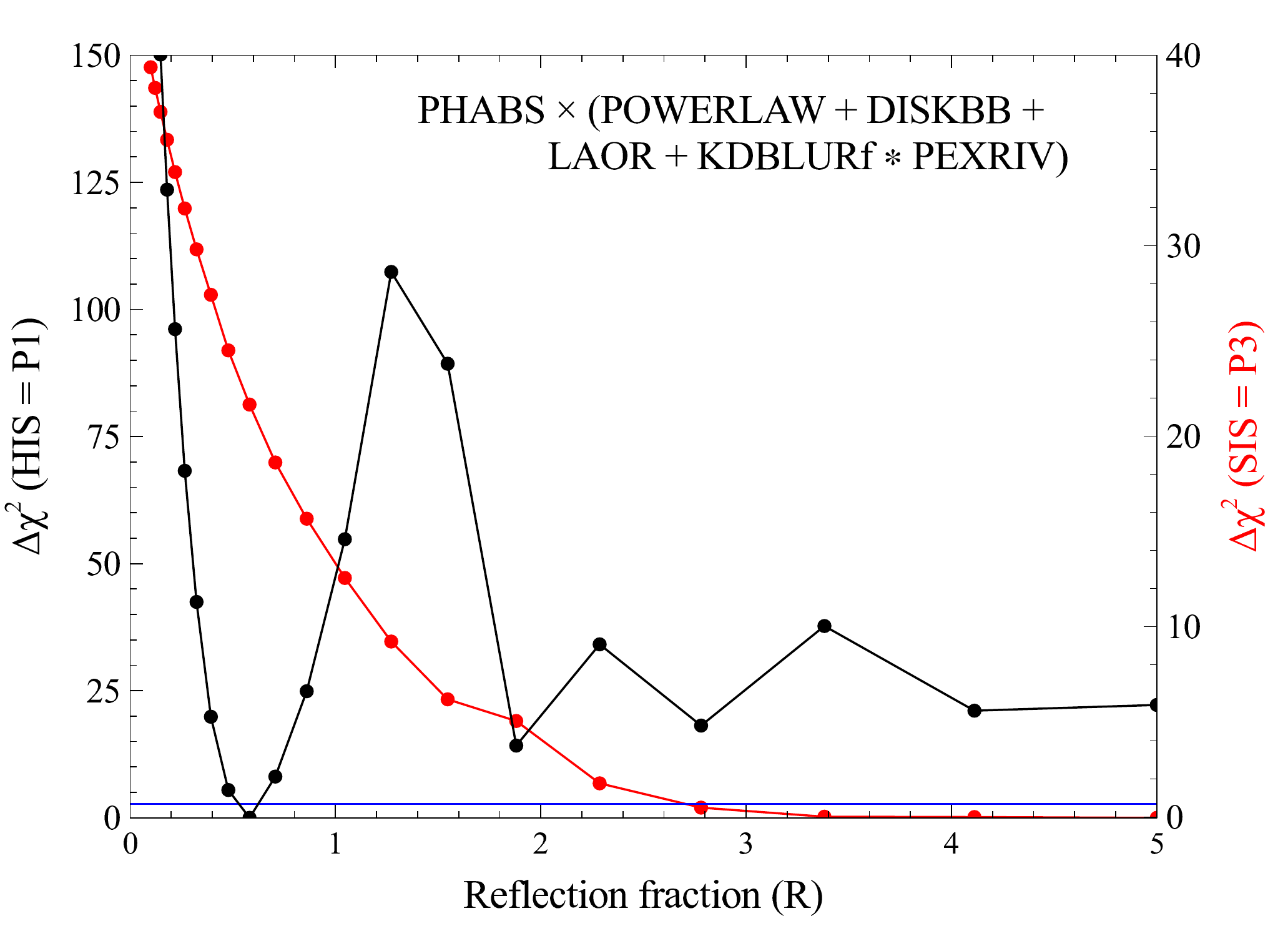} }}}

\caption{Goodness-of-fit versus reflection fraction for the two representative spectra describing the HIS-P1 (black) and SIS-P3 (red). The spectra used refers to those highlighted in Figs. 2 and 4 and the \reflionx\ component inherent in the base model has been replaced with a combination of \pexriv\ together with \laor. The solid blue horizontal line shows the 90~per~cent confidence range. }  
\label{fig5}
\end{figure}

\begin{figure*}[]
\vspace*{0.0cm}
\centering
{\hspace{-0.0cm} \rotatebox{0}{{\includegraphics[width=14cm]{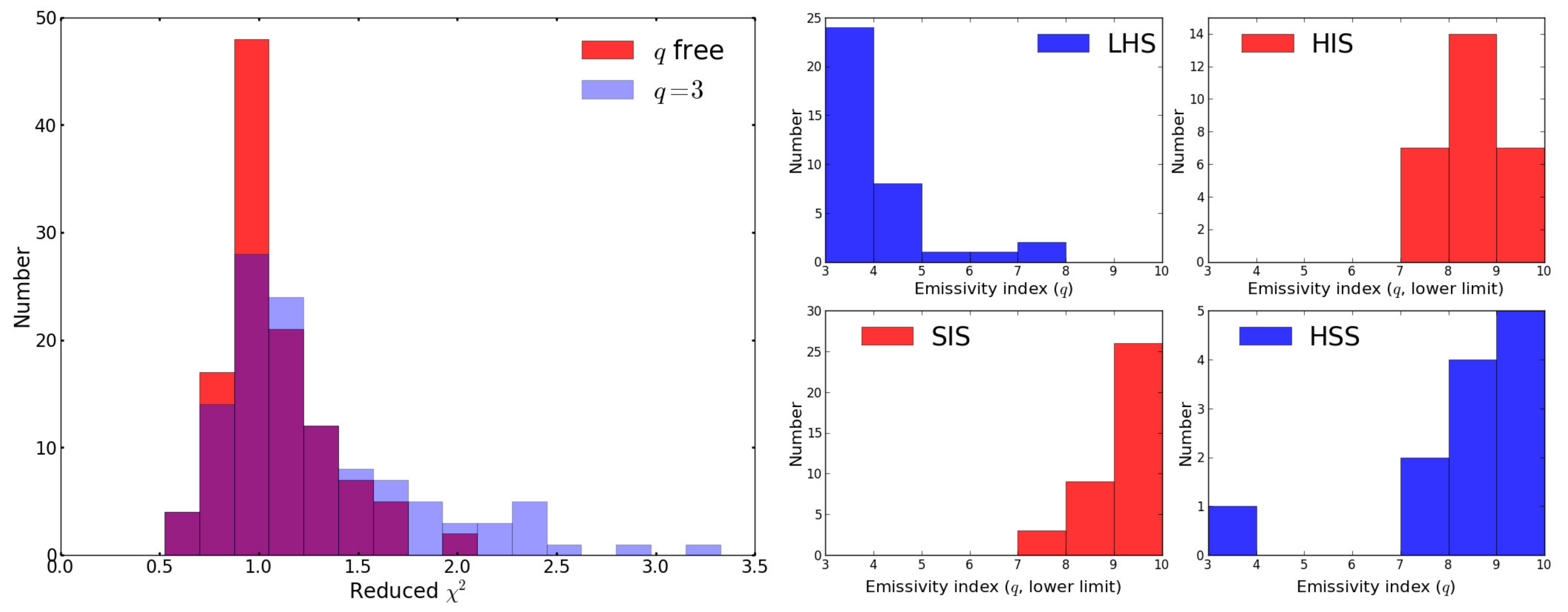} }}}
\vspace*{-0.1cm}

\caption{Left: Distribution of reduced $\chi^2$ for model with a Newtonian emissivity profile ($q=3)$ and a steeper values $\geq 3$. In both cases we see a peak at reduced $\chi^2 =1$, however this is much clearer after allowing the emissivity to vary beyond its Newtonian value. Right: Distribution of emissivity parameter (blue) and their 90\% lower limit (red) for the four states highlighted in each panel Bottom:  In all cases, the emissivity index was constrained to $3\leq q \leq 10$.   }  \vspace*{0.2cm}
\label{fig6}
\end{figure*}

The fact that $\xi$ is maximal in the HSS (P4) despite the fact that this is when the  irradiation appears to be at its lowest level (second panel from the top) can be explained with a number of scenarios. We showed in \S~1.1 that in stellar mass black holes the intrinsic hot disk can result in significant thermal ionization \citep{refbhb} which will be strongest in the disk dominated HSS. Thus, in this scenario, the high ionization measured could also be in part due to thermal ionization.  If this thermal component is the sole source of ionisation in the HSS, $R$ would indeed go to zero.   A further possibility is that the disk is indeed highly photo-ionised as a result of strong focusing of the coronal photons onto the disk. This  would significantly remove the number of hard-photons escaping the system (thus explaining  the second panel from the  top) and cause the disk to be highly ionised. Indeed, observations of disk winds originating in the HSS of various BHBs consistently show winds having  $\xi \sim 10^4 \ergcmps$ \citep[e.g.][]{Miller2008j1655wind,NeilsenLee2009,KingMiller2012,Ponti2012diskjet}. Unfortunately, the lack of reliable constrains on the reflection fraction in the HSS prevents us from making any solid claims on the nature of the disk-corona interaction in this state.

\subsection{Disk Emissivity}

In Fig.~6  (left) we show the distribution  of the reduced $\chisq$ (for 115 degrees of freedom) from all observations assuming the simple Newtonian  `lamp-post'  like geometry in which the  the emissivity profile follows a $q=3$ power-law (light blue),  as well as after relaxing this assumption (red). In both cases there is a clear peak at reduced-$\chisq = 1$, however this peak is much more distinct upon relaxing the Newtonian approximation. The Newtonian approximation naturally does not take into account the effects of general relativity that will  be experienced by the emission from the corona and accretion disk near the black hole. Relativistic effects (strong gravity as well as relativistic time dilation) acts to steepen the emissivity in the inner regions of the disk.  The right panels of Fig.~6 show either the distribution of the emissivity index (blue) or the 90~per~cent lower limit  in their value (red) for the spectral states indicated in each panel.  We refer the reader to the work of \citet{Miniu04, wilkins2011, wilkins2012, fabian2012cyg} and references therein for a detailed study examining non-Newtonian values for the emissivity index, but note here that steep emissivities similar to those found here for the HIS and SIS are a natural and unavoidable consequence of strong gravity.

The bottom panel of Fig.~6 is used here as a simple illustration of the evolution in $q$ as well as the count-rate in both the PCA and HEXTE data. It is clear that $q$ can only be constrained when the PCA data is at its highest level as this constraint does not come from energies $>25$\kev. At the end of the outburst, when the PCA signal-to-noise level drops significantly, the data cannot differentiate between a Newtonian   $q=3$ and a  steeper value.

Following the recipe provided for Cygnus X-1 by \citet{fabian2012cyg} in dealing  with sources where the spin is expected to be high (as is the case for \j), we have repeated our fits with a double emissivity profile such that within a break radius (initially frozen at 4\rg\ but later allowed to vary) the emissivity is $>3$ and beyond it is frozen at 3. The initial value of 4\rg\ for the break radius was chosen based upon the value for Cygnus X-1  \citep{fabian2012cyg}.  We find that, as long as the emissivity is not fixed at  $q=3$, the quality of the fits, and distribution of reduced $\chisq$ are similar to that of a single, unbroken emissivity, and we proceed by using this single power-law emissivity profile as our standard but emphasise that the results presented here do not change if we employ a broken emissivity profile. This is very likely to be due to the comparatively low spectral resolution afforded by RXTE which does not reflect the subtle changes in the reflection profile in a similar manner as \xmm\ or \suzaku\ observations.

\begin{figure*}[!t]
\vspace*{-0cm}
\centering
{\hspace{-0.0cm} \rotatebox{0}{{\includegraphics[width=14.6cm]{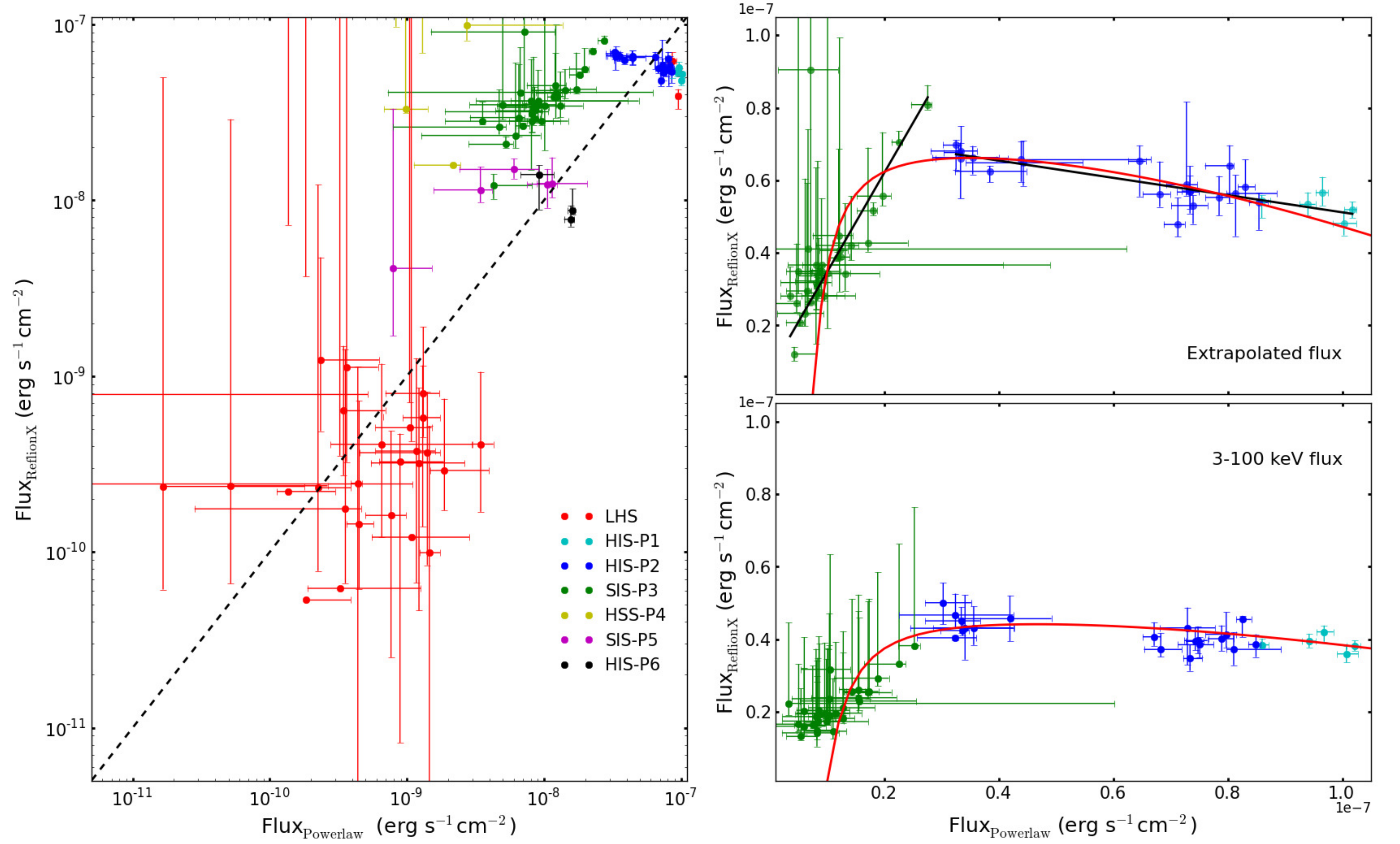} }}}
\vspace*{-0.0cm}
\caption{ Left: Flux-Flux relation during the outburst. The dashed line shows the one-to-one relation. Top-right: Close up of the HIS (blue and cyan) and SIS (green).  The solid black lines shows the best linear fit for each state having a slope of $2.7\pm0.1$ and $-0.24\pm0.02$ in the SIS and HIS respectively. The solid red curve shows the expected flux-flux relation under the light bending model of \citet[][see their Figure 2]{Miniu04} for a system with 60\deg\ inclination, somewhat similar to \j, with a corona varying in height from 1--20\rg. Note that this is not a fit and has been rescaled from the original (see \S~4.3). Bottom-right: Similar but for fluxes between 3--100\kev.  }  
\label{fig7}
\end{figure*}

\begin{table*}
\vspace{-0.4cm}
\caption{Summary of  Spearman's Rank Correlations and Partial Correlation tests on a combination of various model parameters. }  
\centering
\begin{tabular}{cc|cccc|cccc}
\hline
\hline 
• & • & \multicolumn{4}{c|}{HIS (P1 + P2)} & \multicolumn{4}{c}{SIS (P3)}\\ 
\hline 
• & • & \multicolumn{2}{c|}{Spearman's}   & \multicolumn{2}{c|}{Partial } & \multicolumn{2}{c|}{Spearman's}   & \multicolumn{2}{c}{Partial } \\ 
• & • & \multicolumn{2}{c|}{rank-order}   & \multicolumn{2}{c|}{correlation } & \multicolumn{2}{c|}{rank-order}   & \multicolumn{2}{c}{correlation } \\ 
\hline 
Parameter 1 & Parameter 2 & $\rho$ &  p-value & $\rho$ & p-value  & $\rho$ & p-value  & $\rho$ & p-value  \\ 
\hline 

$\xi$ & $F_{\rm powerlaw}$ & -0.953 & $ 2.0\times 10^{-13}$ & -0.473 &  0.014 & 0.319 & 0.071 & 0.092& 0.621 \\ 

$\xi$ & $F_{\rm disk}$        & 0.956 & $ 9.3\times 10^{-14}$ & 0.515 &0.006 & 0.006 & 0.972 & 0.082& 0.656 \\ 

$F_{\rm disk}$ & $F_{\rm powerlaw}$ & -0.955 &$ 1.1\times 10^{-13}$ &-0.427 & 0.0031 & -0.310 & 0.079 &  -0.334 & 0.056 \\ 

$\xi$ & $F_{\rm reflionX}$ & 0.742 & $ 2.2\times 10^{-5}$ & -0.140 & 0.516 & 0.382 & 0.028 & 0.218 & 0.230\\ 

$F_{\rm disk}$ & $F_{\rm reflionX}$ & 0.766 & $ 7.8\times 10^{-6}$ & 0.088 & 0.685 & -0.116 &0.520  &0.140 &0.446 \\ 

$F_{\rm powerlaw}$ & $F_{\rm reflionX}$ & -0.799 & $ 1.7\times 10^{-6}$ & -0.373 & 0.066 & 0.719 & $2.4\times 10^{-6}$ & 0.683 &  $ 4.6\times 10^{-7}$ \\ 
\hline 
\end{tabular} 

\vspace{0.2cm}
Notes:-  Spearman's rank correlations and Partial Correlation test were made for combinations of the reflection, power-law and disk fluxes as well as the ionization parameter. The Partial Correlation test measures the degree of associations between the two parameters listed on the first two columns whilst controlling for the remaining two parameters. The Spearman's coefficient $\rho$ is a measure of the degree of correlation with +1 or -1 indicating a perfect monotone function and 0 a lack of correlation.\vspace*{0.2cm}
\end{table*}

\subsection{Light-Bending and General Relativity} 
   
Hints of  the expected effects of light bending, as described in the introduction, can be seen in the top four panels in Fig.~3.  Most important, is the apparent constancy of the reflection flux in comparison with that of the direct power-law early in the outburst. We investigate this further in Fig.~7. The left panel shows the flux-flux relation  through the whole outburst with the various spectral states shown in different colors. The top-right panel is a close up of the period covering the  HIS and SIS during the first $\sim$~30 days of the outburst\footnote{Excluding the first 3 days when \j\ was in the rising LHS (see Fig.~1).}. Figure~7 is remarkably similar to figure~3 of  \citet{Rossi2005j1650}, where the authors used the flux in the iron line as a proxy for the total reflection in \j. We have superimposed in this figure the expectation from the light bending model for a compact corona varying in height from 1--20\rg\ with a disk having an inclination of 60\deg, from \citet[][see their Figure 2]{Miniu04}. In order to correctly describe the shape of the function shown graphically in \citet{Miniu04}, we used the  Dexter Java application of \citet{dexter_ads} to obtain a fourth-order polynomial fit to their curve from which we applied a linear normalisation of $1.5\times10^{-9}$ and $4.5\times10^{-9}$ to their Y (arbitrary Fe line flux) and X-axis (arbitrary powerlaw flux) respectively.  The model reproduces extremely well the broad  shape of the relation. Finally, the bottom right panel shows this behavior when the non-extrapolated, 3-100\kev\ fluxes are used instead\footnote{In this case the normalisations were $1.5\times10^{-9}$ and $4.5\times10^{-9}$ for the Fe-line and powerlaw flux, respectively). }. We again see that qualitatively, the behavior is the same as above.

As discussed in the introduction, the light bending model of Miniutti et al.~(2004) predicts the existence of semi-distinct regimes in this flux-flux relation. When the corona is located at a height of  $\sim 10$\rg\ the model predicts a flattening in the relation similar to that observed for the HIS (both P1 and P2). The fluxes in this hard intermediate state are clearly correlated, and a Spearman's rank correlation test gives a coefficient of $\rho = -0.799$ corresponding to  a $1.7\times10^{-6}$ chance of a false correlation (Table~1).  The slope of this relation is $-0.24\pm0.02$ (standard error, s.d.)  and this linear fit is shown in the top-right panel of Fig.~7 as a solid black  line. 

As the location of the corona reaches the more extreme environment  within a few \rg\ from the black hole, the model predicts a steep, positive linear relation between the reflection and power-law flux similar to that seen in the SIS, although there is large scatter dominated by poor statistics.  A Spearman's rank correlation test here gives $\rho = 0.719$, with a false correlation probability of just $2.4\times10^{-6}$. The slope of this relationship is  greater than unity, with a best-fit value of $2.7\pm0.1$ (shown as a further solid black  line in Fig.~7) . Note that this is highly inconsistent with the expectation for a static Newtonian corona  with intrinsically varying luminosity, where the slope should be unity across the entire flux range.   It is the combination of a slope~$\gg1$  at low powerlaw flux together with a  near-flattening at higher fluxes that provides evidence for relativistic effects in this case.

It has been suggested \citep[e.g.][]{BallantyneVaughan2003, Ballantyne2011} that changes in the ionization of the inner regions of the accretion disk can give rise to changes the reflection spectrum that can mimic somewhat this flat behavior. In order to robustly assess the strength of the correlations seen in Fig.~7 in both the HIS and SIS, we have performed a number of correlation tests which are summarised in Table. 1.  We performed, for both HIS and SIS,  Spearman rank-order tests for a combination of four parameters (power-law, reflection and disk fluxes as well as the disk ionization) as well as Partial Correlation tests (PCT) for two parameters while controlling for the third and fourth variable. The PCT is of particular importance for our purpose as it removes any potential association  of the ionization parameter (or any other potential source of unwanted correlation) in the flux-flux relations shown in Fig.~7. 

From the  Spearman rank-order tests performed in the HIS (Table~1), it would initially appear  that all four variables are strongly correlated with one another in some way, as all combinations display $|\rho|\gtrsim0.7$. However, after performing the partial correlation test for all combinations we see that for  two of the  previous strong correlations ($\xi-F_{\rm reflionX}$ and $F_{\rm disk}-F_{\rm reflionX}$)  were in fact driven by the  mutual dependence of these parameters on $F_{\rm powelaw}$. These tests clearly indicate that the reflection flux in both states is better correlated with the power-law flux than with ionization and the slope of the correlations (mildly negative in the HIS and strongly positive in the SIS) are highly indicative of  gravitational light-bending in the General Relativistic~regime.

The behavior seen here is consistent with a drop in the height of the corona during the hard-intermediate phase (P1 and P2) followed by intrinsic variations in its luminosity by a factor of a few during the soft-intermediate state (P3). Following the disk-dominated soft-state (P4), the height of the corona increases again (P5 and P6) and the outburst finishes with the intrinsic power dropping as the source goes back into the LHS.

\subsection{$R-\Gamma$ relation}

A strong correlation has been shown to exist between the amplitude of the reflected component ($R$) and the photon index ($\Gamma$)  of the Comptonized spectrum in XRBs {\it in the hard state}  \citep[e.g.][]{Ueda1994rgamma}. This $R-\Gamma$ relation has since been robustly tested by a number of authors \citep[e.g.][]{Gilfanov1999rgamma, Zdziarski1999, Zdziarsk2003, NowakWilmsDove2002gx, ibragimov05} and it is now thought to also apply to Seyferts and radio galaxies \citep[e.g.][]{Zdziarski1999},  further cementing the similarities in the coronal properties at all scales. If this relation indeed turns out to be real (and evidence attests  to this; but see \citealt{Molina2009rgamma}) then this is could be telling us about the  feedback between the hot corona and cold gas in an accretion disk.

We briefly investigate this relationship for \j\ in Fig.~8. For clarity, we only consider data with a fractional uncertainty of less than 50~per~cent. The black arrows approximately show the evolution of the system in time. Although as a whole the data does not strongly support the presence of a relationship between $R$ and $\Gamma$, when the states are roughly separated (different colors) it does appear that early in the outburst through to the  last few days of the hard-intermediate state the relation seems to hold. It is clear, at least, that the rising-LHS and the HIS populate different regions in the figure to the SIS.

The potential presence of  this relation early in the outburst  suggests a  feedback process between the soft photons in the disk and the corona. There are a number of theoretical interpretations for the presence of this correlation \citep{PoutanenKrolik1997,Gilfanov1999rgamma,Gilfanov2000rgamma, beloborodov99, MalzacBeloborodov2001} with the two leading contenders often described as the disk-truncation and dynamic corona model \citep[see][for a detailed study of these models]{done2002review,Beloborodov1999review}. To summarise, in the former, an increase in the reflection fraction is caused by the accretion disk penetrating deeper into a central hot corona thus receiving more illuminating hard photons. The presence of the disk in return offers more soft photons consequently cooling the plasma. For a purely thermal distribution of electrons, the greater the number of soft seed photons, the softer the power-law spectra.  The latter model by \cite{beloborodov99} invokes bulk-motion of the corona above the accretion disk. If the corona is outflowing with mildly relativistic velocities, this would reduce the amount of hard photons hitting the disk, which in turn reduces $R$ and subsequently softens $\Gamma$ as few reprocessed soft photons reach the outflow.  Recent evidence for the coronal plasma ejection model has come from a strong positive correlation between reflected X-ray flux and radio flux in the  black hole binary   Cygnus X-1 \citep{miller2012cyg}.

We will show in \S~4.6 that all evidence points toward the disk radius remaining stable during the HIS to SIS transition (see Figs.~11,12). This constancy effectively rules out the ``disk-truncation" explanation for the $R-\Gamma$ correlation. Instead, one may hypothesise whether the ``outflowing corona" and light bending can be combined to explain the behavior so far detailed for \j.  In the previous section we showed that the flux-flux behavior (Fig.~7) could be explained if early in the outburst (during the HIS) the corona was located relatively far ($\sim10$\rg)  from the black hole and thus  behaved according to  Regime 2 of \citet[][see also \S1.2]{Miniu04}. As the system evolved, the corona  collapsed to  a few~\rg\ and began to experience a higher level of light bending towards the disk (regime~1).  In this scenario, a potential gradient in the outflow velocity of the corona as a function of height could also explain  the behavior seen in Fig.~8. i.e. as the corona collapses from a large height (large outflow velocity; low $R$ and hard $\Gamma$) the outflow velocity decreases ($R$ increases and $\Gamma$ softens) until it becomes effectively static and the system transitions into the SIS.  We will expand upon this possible scenario in what follows and summarise our ideas in \S~5.

\begin{figure}[!t]
\hspace*{-0.5cm}
\centering
{\rotatebox{0}{{\includegraphics[width=9.2cm]{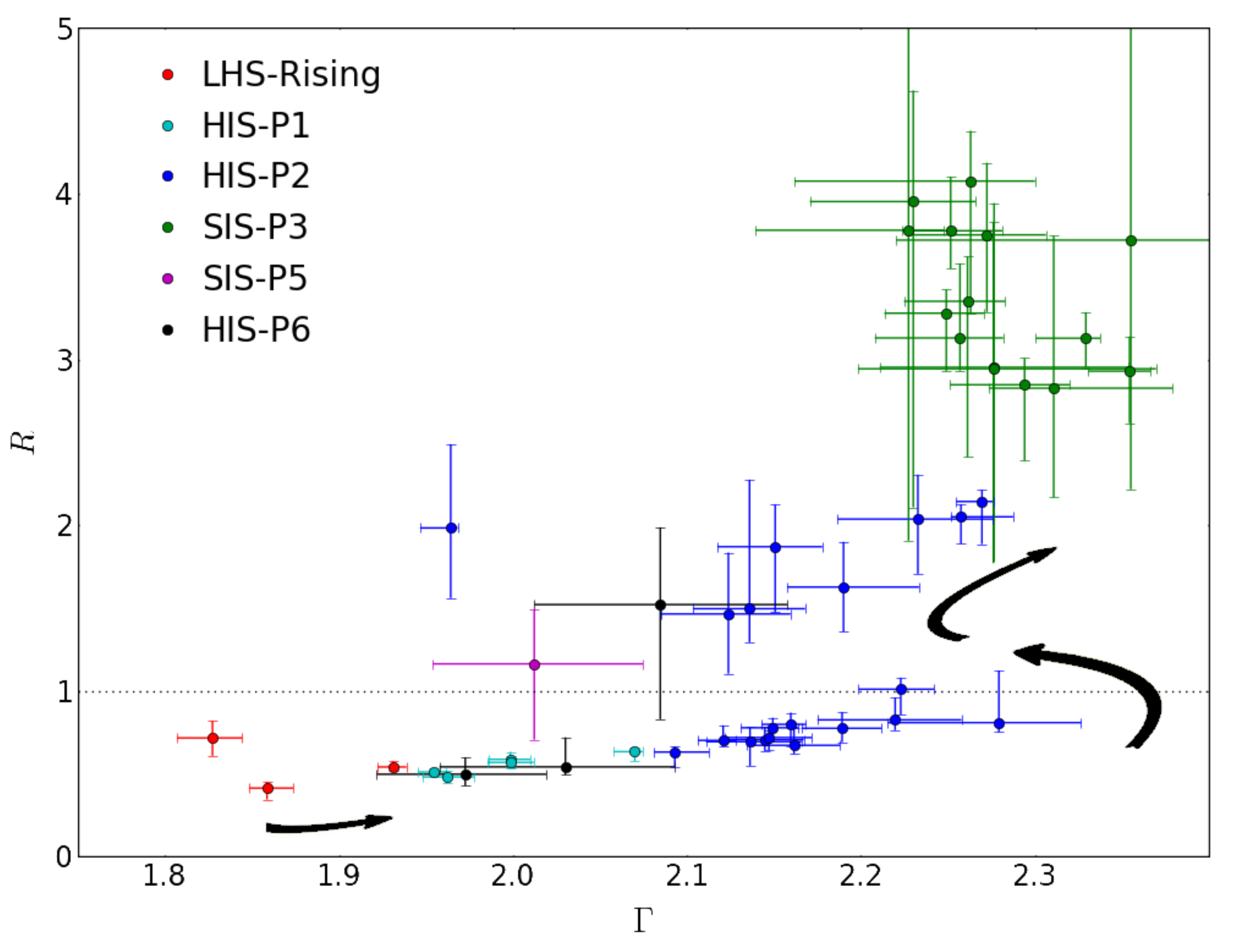} }}}
\vspace{-0.5cm}
\caption{ The well know R-gamma relation appears present in the rising-LHS and during most of the HIS. However, just prior to the transition to the SIS and thereafter, this relation does not hold.  We discuss a potential explanation for this behavior in \S~4.4. Only data with errors less than 50\% their value are used in this figure. The black arrows show approximately the evolution of the system in time. }  \vspace*{0.2cm}  
\label{fig8}
\end{figure}

\subsection{QPOs and Spectral States: A collapsing corona}

 \begin{figure*}[!t]
\hspace*{-0.35cm}
\centering
{\rotatebox{0}{{\includegraphics[width=16cm]{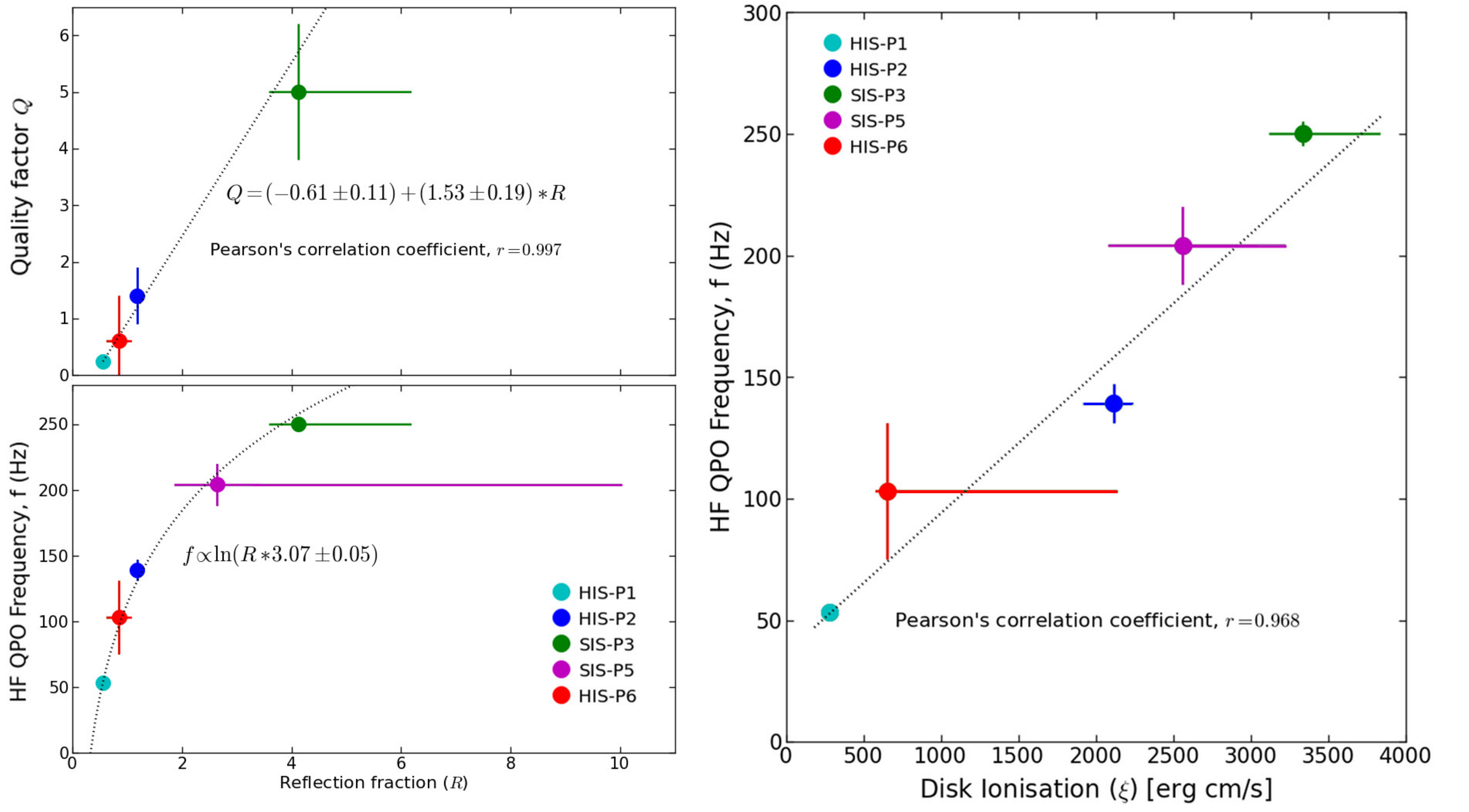} }}}
\vspace*{-0.cm}
\caption{Top-left: HF QPO coherence as a function of reflection fraction.  The relation is well described by a linear function similar to that shown in the figure. Bottom-left: QPO frequency as a function of reflection fraction. The dashed line shows a relation, $f ({\rm Hz}) = (102\pm2)\times {\rm ln} [R \times (3.07\pm 0.05)]$, that fits these data. Right: QPO frequency as a function of disk surface ionisation. These figures show a clear link between the coherence/frequency of the QPO and the reflection fraction or level of disk ionisation,  which we subsequently  interpret as being linked to the size/position of the corona (see \S~4.5). }  
\label{fig9}
\end{figure*}

Throughout this work, we followed the selection made by Homan et al.~(2003) which roughly separates the outburst into six periods coinciding  with significant changes in both the hardness intensity diagram (Fig.~2) as well as in the shape of their power spectra\footnote{In keeping with that work, the HIS is divided into two periods (P1 and P2). Here, we also add two extra periods which we have denoted as the LHS-rising and falling.} As highlighted in the introduction (\S1.3), those authors demonstrate the presence of high-frequency (HF) variability in \j, together with a  HFQPO which was shown to evolve in both frequency and coherence during the outburst.  The highest frequency which was reliably measured was at  $\sim250$Hz in the SIS, with the frequency being much lower ($\sim50$Hz) at the onset of the outburst.

In Fig.~9 (top-left), we show the presence of a strong (Pearson's $r=0.997$) positive relation between the reflection fraction and quality factor of the QPO ($Q-R$ relation). In order to create this figure, we have averaged the values of $R$ shown in Fig.~3  for each of the periods in question and used the values for the coherence provided by Homan et al.~(2003; Table~1). The bottom-left panel shows the frequency of the HF QPO as a function of $R$ ($f-R$ relation).  We also show in Fig.~9 (right), the QPO frequency as a function of disk (surface) ionisation parameter.

QPOs are notoriously difficult to explain and it is not our purpose to provide a  quantitative description of this phenomenon. However, it is worth stressing that most models \citep[e.g.][]{Nowak1997, CuiZhangChen1998,  PsaltisBellonivanderKlis1999, StellaVietriMorsink1999, McKinneyqpo2012} strongly link the origin of QPOs with orbits and/or resonances in the inner accretion disk close to the black hole. Current models cannot fully explain, in a physical manner,  the range in coherence observed in various systems nor the manner in which the frequencies change with states. 

To explain the  range in coherence observed in accreting neutron stars, \citet{BarretOlive2006,BarretOlive2007} devised a toy model in which the  changes in $Q$ are driven by changes in the scale height of the disk. A small scale height gives rise to high coherence and vice versa. Expanding on this idea, it appears that at least for \j, it is physical changes in the radius/size of the corona that give rise to changes in both quality factor $Q$ and QPO frequency. To illustrate this hypothesis, consider  Fig.~9 (bottom) together with our interpretation of the behavior displayed in Fig.~7 (\S~4.3). Early in the outburst the frequency of the HF QPO  appeared at $\sim55$Hz.  The Keplerian frequency at a given radius is $f ({\rm Hz}) \approx 3.2\times 10^4  M^{-1} r^{-3/2}$, where $M$ and $r$ are in units of Solar mass and Gravitational radius respectively. Hence, during the brief P1 period, the HF QPO frequency is close to the orbital frequency at $\sim28$\rg\ assuming a $4\msun$ black hole and potentially moves to $\sim15$\rg\ in the second half of the HIS. As the outburst continues and the corona continues to collapse, it is plausible that the frequency continues to increase (corresponding to $\sim10$\rg\ in the SIS) eventually approaching a value that should be consistent with the  Keplerian frequency at the ISCO. The continued decrease in the size of the corona gives rise to the increase in coherence. In this scenario, the frequency of the QPO should relate to the size of the corona and thus would naturally increase as the corona collapses. The relationship between the QPO frequency and the surface ionisation parameter could be suggestive of an intrinsic relationship between the irradiation of the disk and its  magnetic field properties,  the latter which has recently been proposed as a possible means to  produce(low-frequency) QPOs \citep[see e.g.][]{oneill2011qpo, Oishi2011qpo}. This possibility will be addressed in forthcoming work.

Finally, the relation between the coherence of the HF QPOs and the reflection fraction also leads to the interesting prediction  that HF QPOs should only be observed when $R\gtrsim 0.4$ when the coherence is $Q>0$. This results is consistent with observations, with no HF QPO ever having been found in the LHS where $R\lesssim1$.

\subsection{Radio (jet) Emission and Reflection Fraction}
\label{jet}

 \citet{CorbelFender2004} presented a comprehensive analyses of the radio emission observed  during the outburst of \j. In that work, the authors suggest that the transition between the HIS and SIS\footnote{Referred to as the intermediate and steep powerlaw states respectively in \citet{CorbelFender2004}.}  is associated with a massive radio  ejection event. The observations in the LHS was found to be consistent with the presence of a steady, compact jet, as is often seen in the LHS of black hole binaries \citep[see e.g.][and references therein]{Fender2001jets,Fender091}. This potential ejection event during the HIS--SIS transition coincides with the time where we see a sharp jump in the reflection-fraction. In Fig.~10 we show the radio flux density\footnote{Obtained directly from  Table~1 of \citet{CorbelFender2004}.  We are using the flux densities at 4800~MHz  for all observations, except for the rising-LHS where this was not available. In this case,  we proceeded by averaging the values presented for 1384 and 2496~MHz.}  vs the reflection fraction calculated herein. At the time of the steady compact jet (in the LHS \& HIS), the radio flux density
increased by a factor of $\sim5$ with no statistically significant change in
$R$. However, immediately following the radio ejection in the SIS, the reflection
fraction increases dramatically resulting in a bi-modality in the radio-flux
density--reflection fraction plane. \textit{This suggests an intimate link between the
  jet ejection site and the collapsing corona.}  Indeed, \citet{beloborodov99}  predicts a link between radio jets and reflected flux, which was also   seen in Cygnus X-1 \citep{miller2012cyg}.
 
At later stages in the outburst (HIS-P6, falling LHS), the measured radio flux density
is likely a dominated by emission from the zone where the ejected plasma interacts
with the ambient ISM surrounding the binary system.  Unfortunately, due to the low spatial
resolution of the radio observations, this emission is not resolved from that due to
any reformation of the steady jet close to the black hole.

 \begin{figure}[!t]
\hspace{-0.5cm}
{\rotatebox{0}{{\includegraphics[width=9.cm]{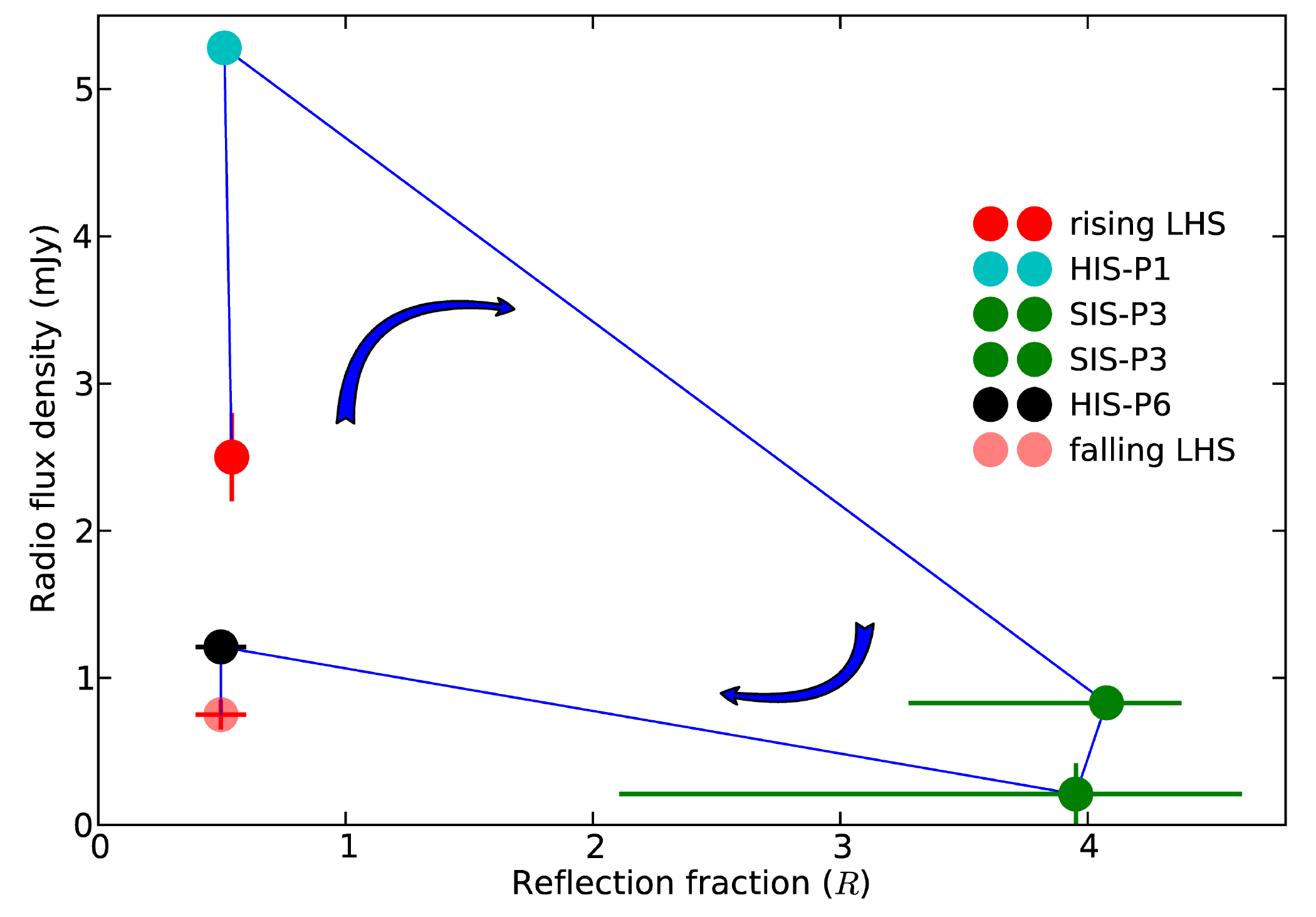} }}}
\vspace{-0.4cm}
\caption{ Radio flux density as a function of reflection fraction. The arrows show the direction of the outburst. There appears to be two distinct branches where the reflection fraction is either constant at $R\sim0.5$ or at $R\sim4$. These two branches corresponds to the LHS/HIS and SIS respectively. Radio data obtained from \citet{CorbelFender2004}.   }\vspace*{0.2cm}  
\label{fig10}
\end{figure}

\label{spin}
\subsection{Disk (inner) Radius and State Transition}

\begin{figure*}[!t]
\vspace*{-0cm}
\centering
{\hspace{-0.0cm} \rotatebox{0}{{\includegraphics[ width=16.5cm]{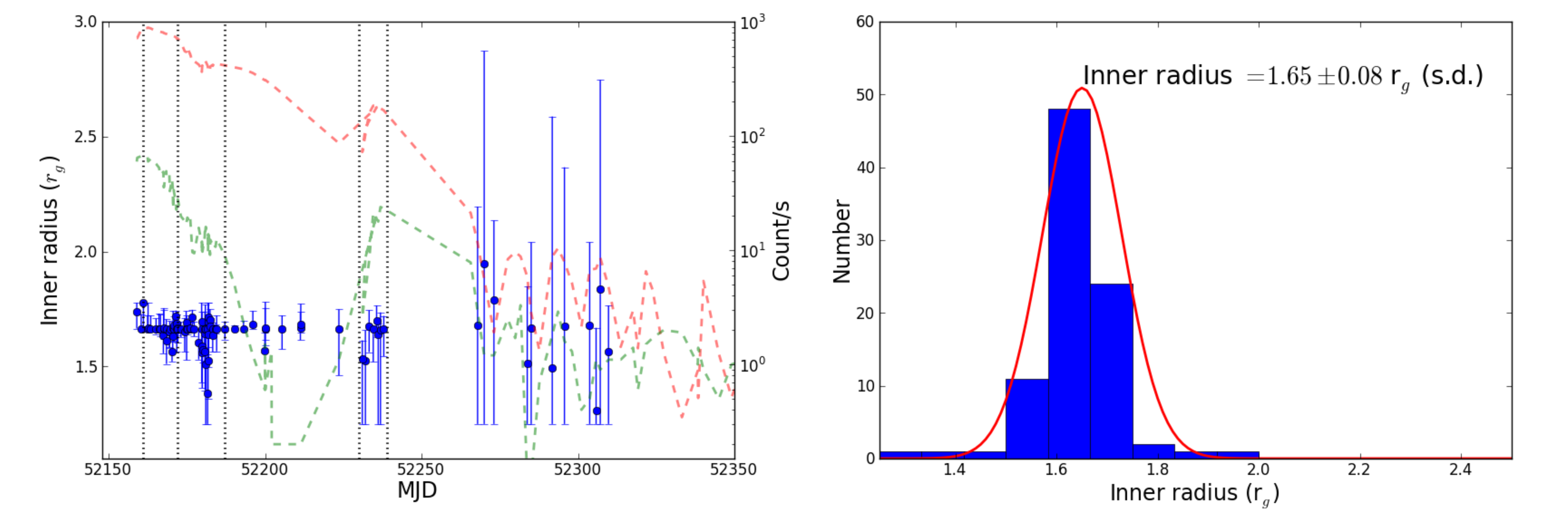} }}}
\vspace*{-0.cm}
\caption{Left: The inner radius during the outburst are shown in blue. We only show measurements where the uncertainty in the radius  is $< 50\%$ of its value. The dashed lines show the evolution of the PCA (grey) and HEXTE-A (green) count rate in a similar manner to Fig.~6 (bottom). Right: Distribution of inner radii having errors $\leq50\%$ its value. In red we show the standard normal distribution with a mean radius of 1.65\rg\ and standard deviation of  0.08\rg. }  
\label{fig11}
\end{figure*}

\begin{figure*}[]
\vspace*{-0.4cm}

\centering
{\hspace{-0.0cm} \rotatebox{0}{{\includegraphics[ width=8.cm]{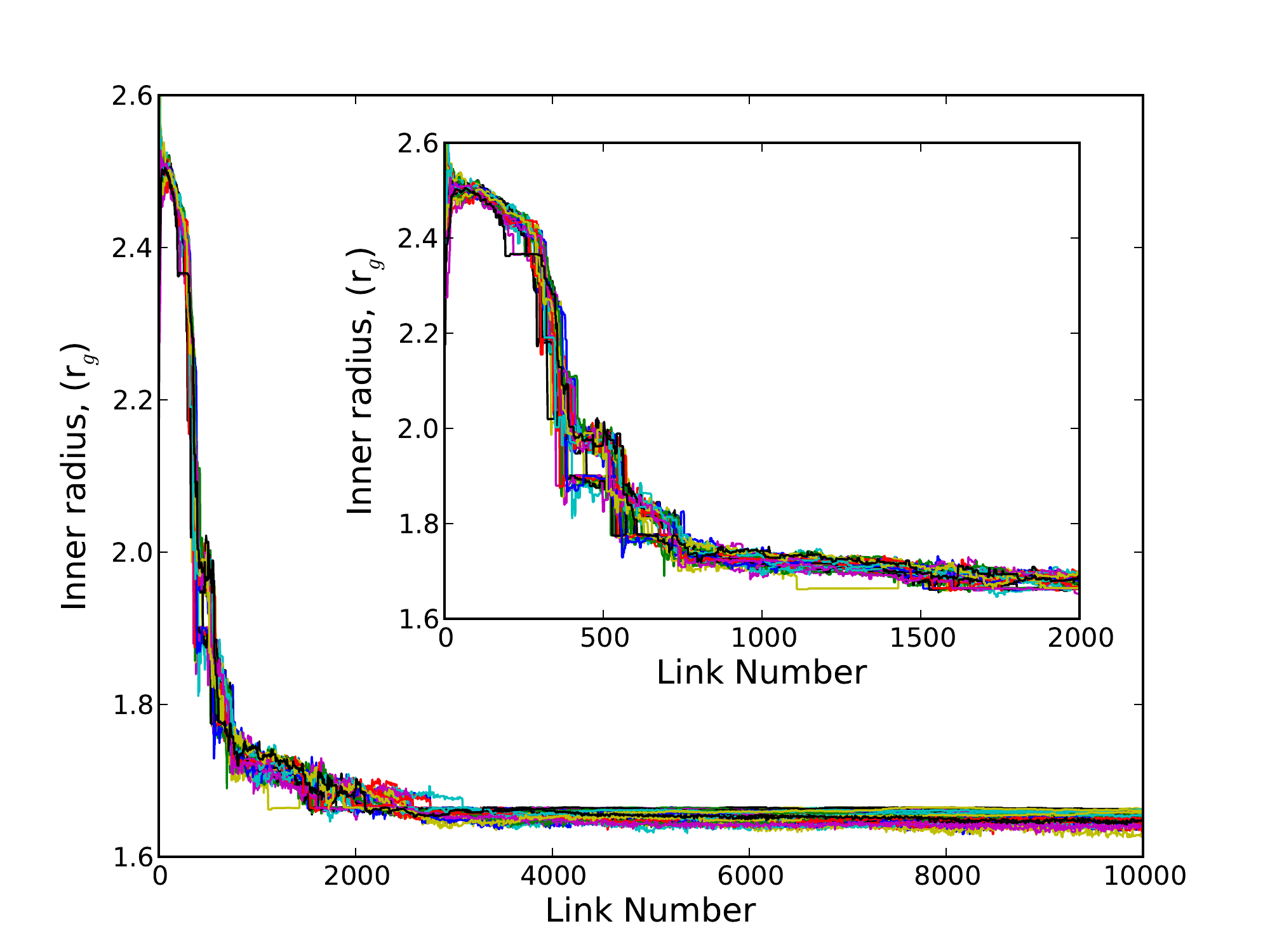} }}}
{\hspace{-0cm} \rotatebox{0}{{\includegraphics[ width=8.cm]{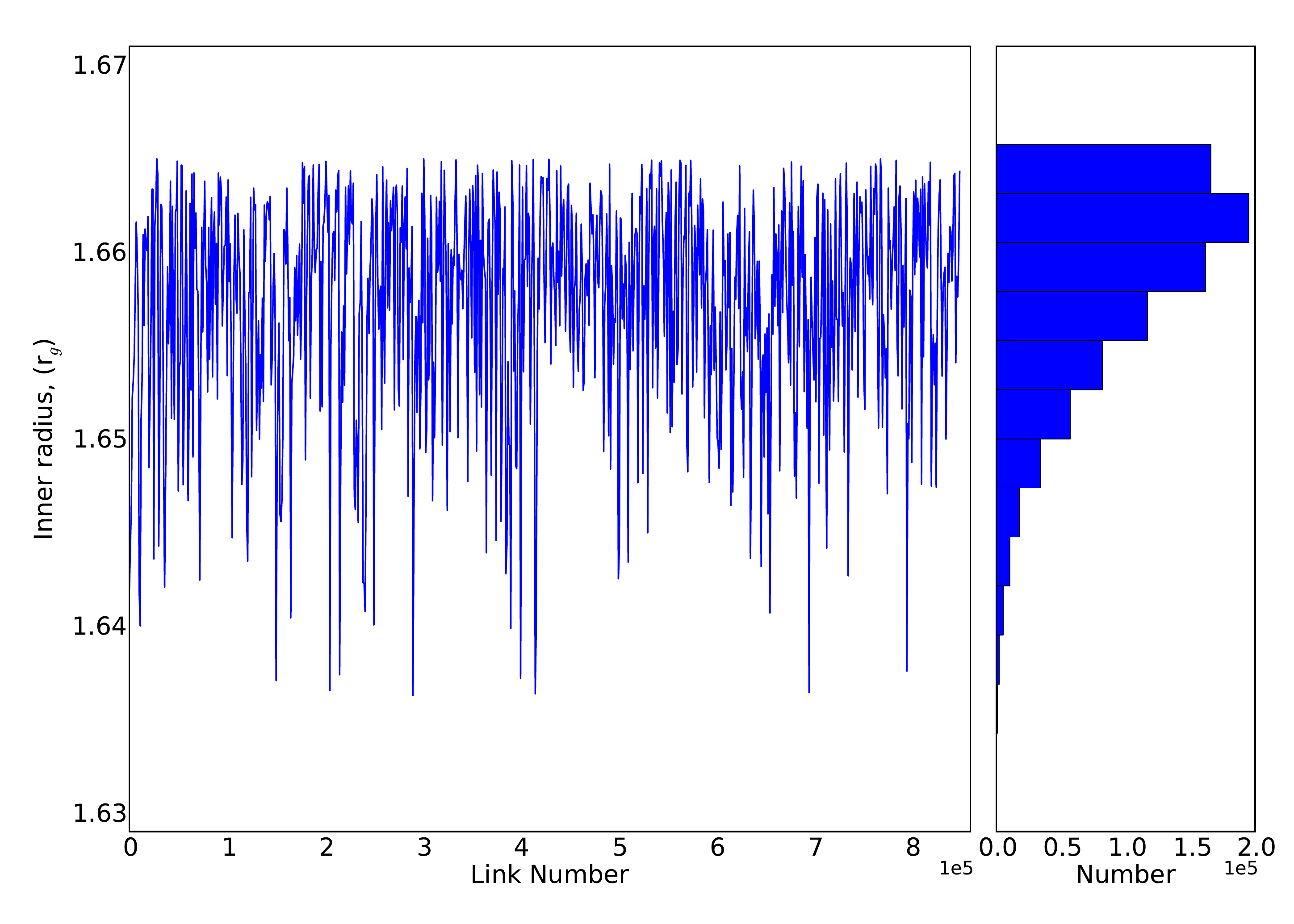} }}}

\vspace*{-0.cm}
\caption{MCMC results for  the inner radius (a proxy for spin) obtained from the simultaneous fit to the first ten spectra in the HIS. Left: Figure tracing 50 out of a total of 170 ``walkers" during their random walk. The figure shows that the various chains converge pretty quickly indicating efficient sampling. The inset shows a close up of the first 20,000. steps.  Right: Full MCMC containing all 170 walkers, after having the first 5,000 elements ``burnt-in". For clarity, we only show every 1000th element of the chain.}  \vspace*{0.2cm}

\label{fig12}
\end{figure*}

As has been discussed throughout this paper, a popular explanation for state transitions is a radial variation in the extent of  the  accretion disk. This model has been highly successful in part due to its flexibility and ease in which it can explain the ``weak" reflection fraction ($R<1$) often found in the LHS. However, over the past few years it has consistently been shown that in the luminous phases of the LHS -- at least above $\sim1\times 10^{-3}~L_{\rm Edd}$\footnote{Contrast this with the broadband analyses of GX~339-4 presented by \citet{tomsick09gx} where the authors find clear evidence for the recession of the accretion disk beginning  only at Eddington luminosities below $\sim\times10^{-3}~L_{\rm Edd}$.  } -- the disk does not appear to be truncated away from the radius of the ISCO \citep{Millergxlhs2006,  MillerHomanMiniutti2006, miller08gx, reisj1118, reislhs, Reynold2011swift, Waltonreis2012}.   Figure~11 (left) shows that the present work can statistically rule out a disk being truncated further than $\sim3$\rg\ even in the LHS. During the brighter, intermediate states, we constrain this radius to $\sim1.65$\rg. This adds support to the idea that the inner disc radius remains roughly constant throughout the  LHS--HIS--SIS state transitions in black hole binaries.

Where we have not been able to constrain the radius, this has largely been due to the data quality (the falling phase of the LHS is inherently less luminous)  as well as the fact that reflection is intrinsically weaker in the LHS. The strongest reflection features are expected in the intermediate states where the disk receives a larger fraction of the hard X-ray emission \citep[see e.g.][]{hiemstra1652}. Note also that despite the comparatively  low spectral resolution afforded by the \rxte -PCA (18\% FWHM energy resolution at 5.9\kev)~\citet{WilmsNowak2006cyg} showed that this instrument  can indeed resolve line-widths down to a least  $\sigma \sim0.3$\kev. Higher resolution observations of \j\ early in the outburst showed that, when modelled with a \gaussian, the Fe\ka\ emission line is consistent with having a width of~$\sigma \sim~1.1$\kev~\citep[Table~4 in][]{Waltonreis2012}\footnote{The original analyses of this \xmm\ dataset performed by \citet{miller02j1650} included an extra smeared edge component at $\sim6.8$\kev\ which resulted in the \gaussian\ having a width of only $\sim250$\ev. }.

Assuming  the stable radius shown in Fig.~11 for the two intermediate states is indeed the radius of the ISCO, we find, using the relationship between ISCO and black hole spin of \citet{Bardeenetal1972}, a dimensionless spin parameter of  $0.977^{+0.006}_{-0.007}$ consistent with the value found in  detailed analyses of single, high quality data obtained with \xmm\ ($0.84\leq a \leq 0.98$; \citealt{Waltonreis2012}) or  \bepposax\ ($a \gtrsim 0.93$; \citealt{MiniuttiFabianMiller2004j1650}\footnote{Spin converted from the lower limit on the inner radius of $\sim2.1$\rg.}).

As a test of the robustness of this result, we have performed a joint fit to the first ten observations in the HIS. We used the same base model as before with each individual observation having their own set of parameters -- disk emissivity index, temperature, normalisation and  parameter,   power-law index, as well as the normalisation of the powerlaw and reflection component. However, this time we  forced the inner radius in the various observations to be a global parameter thus assuming a constant value.

This simultaneous fit contains a total of 81 free parameters\footnote{The sheer number of free parameters and computational time required to do  $\chisq$ fitting as well as the MCMC analyses described in what follows drove the need to constrain this analyses to only 10 observations as opposed to all 116.} and with this comes a high chance of mistaking a local minima in $\chisq$ space for the global best fit. In order to address these limitations, we proceeded by minimised the fit using standard $\chisq$ fitting techniques within \xspec\ until a reasonable fit was produced ($\chisq/\nu <2$) at which point we halted the minimisation\footnote{The actual quality of the fit at this time was $\chisq/\nu = 2059.7/1149$.} and proceeded with Monte Carlo Markov Chain (MCMC) analysis.  We employed the MCMC procedure described in \citet{mcmc2012} (code  found at \href{http://danfm.ca/emcee/}{http://danfm.ca/emcee/}) and implemented in the \xspec  spectral fitting package by Jeremy Sanders (\xspec implementation described in \href{https://github.com/jeremysanders/xspec_emcee}{https://github.com/jeremysanders/xspec\_emcee}). MCMC techniques have been successfully used to address similar problems in constraining the black hole spin of NGC~3783 \citep{Reynolds20123783} as well as in modelling the kinematics of the microquasor XTE~J1550--564 \citep{SteinerMcClintock2012jet1550}.

We added  a 5\% random perturbation to all the parameters in the fit  described above, and increased the value of the  inner radius from the  starting  value of $\sim1.6$\rg\ to 2.5\rg\ in order to guarantee that the chain could freely converge to the global minimum.  We used a total of 170 ``walkers", each iterated (``walking")  10,000 times. Figure~12 (left) shows the evolution of the walk in inner radius for 50 randomly selected walkers. It is clear that the walkers converge to the same value  efficiently. Nonetheless, in order to be conservative we have ignored (``burned-in") the first 5000 elements of each chain and show on the right the full MCMC chain for the radius which is clearly well behaved with a peak distribution at $1.66\pm0.01$\rg\ (s.d.), in excellent  agreement with the results in Fig.~11.

\section{Summary}

In this work, we try to take a broad systematic approach to not only the data reduction but also in the manner in which the spectra are fit during all spectral states observed during the outburst. We have presented a number of empirical results inferred from our a priori assumption that the observed spectra and their evolution is a consequence of variation in three separates emission components -- the power-law continuum, thermal disk and reprocessed reflection.  Although we have tried to convey a fully consistent, albeit qualitative picture of the spectral evolution of \j, it is very difficult to obtain unique interpretation of the results and nearly impossible to generalise this to all systems. In a forthcoming publication, we will apply similar techniques to a larger set of objects at which point we hope to be able to make a stronger statement regarding the global population of stellar mass black hole binaries and possibly AGN. 

For now, we summarise  the main results of the presented spectral analyses as follows:

\begin{enumerate}
\item The outburst is well characterised by a model consisting of a hard X-ray continuum, a thermal disk component and reprocessed emission (reflection).

\item The emissivity profile of the disk is not well characterised by a simple Newtonian approximation and, where this value can be constrained early in the outburst, it is steeper than the Newtonian value $q=3$. 

\item The $F_{\rm reflection}-F_{\rm powerlaw}$ plane for the periods covering the hard-intermediate to the soft-intermediate spectral states displays two distinct 
behaviors. Early in the outburst, during the HIS, this plane is nearly flat (slope of $-0.24\pm0.02$)  with the powerlaw flux 
varying by $\sim 5\times$ and the corresponding reflection flux by $\sim1.5\times$ During the  SIS that followed, this behavior become distinctively different with a $F_{\rm reflection}-F_{\rm powerlaw}$ slope of $2.7\pm0.1$.

\item  The nearly-flat behavior of the $F_{\rm reflection}-F_{\rm powerlaw}$ plane seen for the HIS cannot be explained away by variations in the ionization of the accretion disk and we propose that the most likely explanation is that of the the light bending model of \cite{Miniu04}.

\item The  HIS--SIS transition is accompanied by a sharp increase in the reflection fraction, which we interpret  as a sudden collapse of the corona as the system approaches the thermal state. The collapsed  corona now experiences stronger effects of light-bending due to its proximity to the black hole. The radiation from the corona is focused towards the disk thus systematically increasing the reflection and decreasing the continuum.

\item We confirm the  $R-\Gamma$ correlation during the low-hard and early hard-intermediate states but find that this relation does not hold once the system has transited into the SIS.

\item We find a strong linear correlation between the reflection fraction and the coherence of the high frequency QPOs. We also find a correlation between the frequency of the QPO and $R$ which is well explained with a simple log-linear relation.

\item We find a strong correlation between the frequency of the QPO and the ionisation state of the accretion disk.  This relationship is suggestive of an intrinsic relationship between the irradiation of the disk and its magnetic field properties.

\item We have presented a scenario within the collapsing corona toy model  where the increase in the coherence of  QPO  with increasing reflection fraction is a consequence of decreasing emitting region as the corona collapse to regions closer to the black hole where light bending significantly increases $R$. Similarly, as the corona collapses, the increase in the  QPO frequency  could be associated with a decrease in the radial extent of the corona.

\item In the low-hard and hard-intermediate states, the radio flux
  density varies by a factor of $\sim5$ with no change in the reflection fraction. The
  ``ballistic" radio ejection associated with the HIS--SIS transition is accompanied
  by a sharp increase in the reflection fraction.

\item We show that the HIS-SIS transition is not due to variations in the inner radius of the accretion disk which is found to be stable at $1.65\pm0.08$\rg. Assuming this is the radius of the innermost stable circular orbit, we find a spin of $a\gtrsim0.96$ in excellent agreement with values measured in other works.

\item Our work shows that the configuration of the corona  -- and possibly that of the  magnetic field -- is instrumental in defining the state of the system.

\end{enumerate}

\section*{Acknowledgements}
RCR thanks the Michigan Society of Fellows and NASA.  RCR is supported by NASA through the Einstein Fellowship Program, grant number PF1-120087 and is a member of the Michigan Society of Fellows. ACF thanks the Royal Society. 

\vspace*{0.5cm}

\bibliographystyle{mnras}

\bibliography{bibtex.bib}

\appendix

\section{Appendix material}

\end{document}